\documentclass[prd,aps,nofootinbib,floatfix,10pt]{revtex4}
\usepackage{graphicx,epsfig,mathtools}
\usepackage[usenames]{color}
\usepackage{multirow}
\usepackage{slashed}
\usepackage{epstopdf}

\begin{document}
\title{Weak decays of triply heavy baryons in light front approach}

\author{  Wei Wang$^1$,  Zhi-Peng Xing$^2$\footnote{Email:zpxing@sjtu.edu.cn}}
\affiliation{$^{1}$
INPAC, Key Laboratory for Particle Astrophysics and Cosmology (MOE), Shanghai Key Laboratory for Particle Physics and Cosmology, School of Physics and Astronomy, Shanghai Jiao Tong University, Shanghai 200240, China }		
\affiliation{$^{2}$  Tsung-Dao Lee Institute, Shanghai Jiao Tong University, Shanghai 200240, China}

\begin{abstract}
We analyze  weak decays of triply heavy baryon $\Omega_{ccc}^{++}$ and calculate the  form factors for $\Omega_{ccc}^{++}\to \Xi_{cc}^{++}/\Omega_{cc}^+$ transitions in a light front quark model. 
The momentum distributions in form factors are accessed via the pole-model parametrization.  Using the results for form factors,  we  predict decay branching fractions for semi-leptonic decays, and obtain $\mathcal{B}(\Omega_{ccc}^{++}\to\Xi_{cc}^+e^+\nu_e)=0.08\%$,  $\mathcal{B}(\Omega_{ccc}^{++}\to\Omega_{cc}^+e^+\nu_e)=1.23\%$.   
Forward-backward asymmetry and angular distributions are then investigated  in this work. We also estimate   branching fractions for  a few  color-allowed processes and find that the $\Omega_{ccc}^{++}\to\Omega_{cc}^+\pi^+$ has a sizable branching fraction:  $\mathcal{B}(\Omega_{ccc}^{++}\to\Omega_{cc}^+\pi^+)=1.82\%$.   We point out that the $\Omega_{ccc}^{++}\to\Omega_{cc}^+\pi^+$ is a golden channel for the discovery of the $\Omega_{ccc}^{++}$. This analysis can provide a useful reference for future experimental search in future.
\end{abstract}

\maketitle

\section{Introduction}

The  quantum  description of 
strong interactions of quarks and gluons is embedded in  gauge field theory of SU(3), namely quantum chromodynamics. While the high energy behavior can be understood with perturbation theory, the low energy property is nonperturbatively  manifested with hadrons. The study of hadron spectroscopy provides a way to decipher the mysterious structure of QCD. 
In the past decades there is a renaissance for the study of hadron spectroscopy. Needless to mention the hadron exotics that defy the standard quark model interpretation~\cite{CDF:2003cab,LHCb:2014zfx,LHCb:2019kea},  many  conventional hadrons are also found on various experimental facilities ~\cite{LHCb:2017iph,Belle:2017ext,LHCb:2018vuc,LHCb:2019bem}. 

\begin{figure}[htbp!]
\includegraphics[width=0.4\columnwidth]{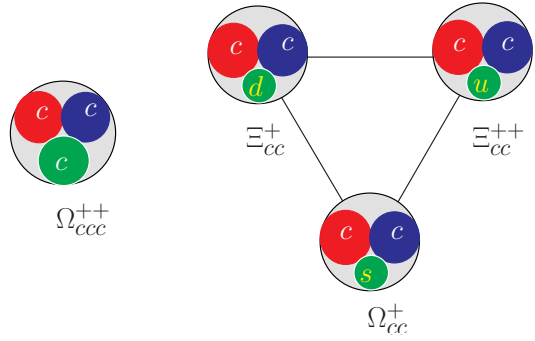}
\caption{Triply charm and  doubly charm baryons.}
\label{fig:doubly_heavy_baron}
\end{figure}

Among various hadrons,  hadrons with one or more heavy quarks are of special interest\cite{Hasenfratz:1980ka,Bjorken:1985ei,Silvestre-Brac:1996myf,Vijande:2004at,Roberts:2007ni,Martynenko:2007je,Guo:2008he}. Since the heavy quark inside is nonrelativisticaly, it provides a static color source, and thus the study of heavy hadrons has received great attentions.
In 2017, the LHCb collaboration has firstly discovered the doubly heavy baryon $\Xi_{cc}^{++}$~\cite{LHCb:2017iph} in the final state~\cite{Yu:2017zst}, which has triggered more interests in hadron physics both on theoretical and experimental sides~\cite{Qin:2021wyh,Hu:2019bqj,Han:2021gkl,Yu:2019lxw}. In addition to the confirmation from $\Xi_{cc}^{++}\to \Xi^+_c\pi^+$ decay mode~\cite{LHCb:2018pcs}, the lifetime has been measured by LHCb collaboration~\cite{LHCb:2018zpl}, and meanwhile a theoretical update of lifetime calculation is also presented in Ref.~\cite{Cheng:2021qpd}.   In 2022, a new decay mode $\Xi_{cc}^{++}\to\Xi_{c}^{+\prime}\pi$ is  observed~\cite{LHCb:2022rpd}.  Thus from this viewpoint the $\Xi_{cc}^{++}$ has been well established on the experimental side.  As a consequence, the last missing category of hadrons is the triply heavy baryon, which is made of three heavy charm or bottom quarks. 
There have been a number of triply heavy baryon studies now\cite{Brambilla:2005yk,Jia:2006gw,Zhang:2009re,Brambilla:2009cd,Meinel:2010pw,Chen:2011mb,Wang:2011ae,Flynn:2011gf,Llanes-Estrada:2011gwu,Aliev:2012tt,Wei:2015gsa,Yang:2019lsg}. In this work we will focus on the $\Omega_{ccc}^{++}$  shown as the left panel of Fig.~\ref{fig:doubly_heavy_baron}, and more particularly we explore the golden final state to reconstruct this hadron.

In a previous analysis ~\cite{Wang:2018utj,Huang:2021jxt}, the flavor SU(3) symmetry is used to analyze various decays of  triply heavy baryons. Though the pertinent decay amplitudes are obtained in a general way and relations for partial widths are derived, it is still lack of explicit predictions for decay branching fractions. Induced by the $c\to d/s$ transition, $\Omega_{ccc}^{++}$ can decay into the $\Xi_{cc}^+ $ or $\Omega_{cc}^+$, which form  an SU(3) triplet shown as the right panel of Fig.~\ref{fig:doubly_heavy_baron}.   On the theoretical side,   decay amplitudes for semileptonic decays are parameterized by form factors.  Quite a few theoretical tools are developed to calculate these form factors, while in this work the light-front quark model (LFQM) will be employed~\cite{Jaus:1999zv,Jaus:1989au,Jaus:1991cy,Cheng:1996if,Cheng:2003sm,Cheng:2004yj}. 
With the simplification  of quark-diquark assumption, baryons are similar with mesonic systems. 
The two spectator quarks in a baryon play the role of the antiquark in a mesonic system and are treated as a system of spin-$0$ or $1$.  Thus this method is widely used for the analysis of baryonic transition form factors and a number of interesting results are obtained~\cite{Ke:2007tg,Wei:2009np,Ke:2012wa,Ke:2017eqo,Zhu:2018jet,Hu:2020mxk,Chang:2020wvs,Chen:2021ywv}. Recently, a three-body vertex function in LFQM has been explored~\cite{Ke:2019smy}, in which the new vertex function is used in $\Lambda_b\to\Lambda_c$ and $\Sigma_b\to\Sigma_c$ decays. It is demonstrated that the two approaches give the consistent results, and thus validated the diquark approximation from certain viewpoint.

The semileptonic decays are governed by the form factors, and thus branching fractions and angular distributions can be explored directly. However due to the low efficiency to reconstruct the $\Omega_{ccc}^{++}$ with the missing neutrino, it is very hard to discover the $\Omega_{ccc}^{++}$ in these semileptonic channels. Towards the identification of the most likely final state to reconstruct the $\Omega_{ccc}^{++}$, we also explore the nonleptonic decay modes induced by the charged  $W^+$ emission. To the end, we find that the $\Omega_{ccc}^{++}\to \Omega_{cc}^+\pi^+$ has a sizable branching fraction. Thus  it is presumable that an experimental search through this channel can lead to the discovery of  $\Omega_{ccc}^{++}$.

The rest of this paper is organized as follows.  In Sec.~II,  the theoretical framework of our calculation is given.  After the parametrization of form factors,  we will present the framework of the light-front approach under the diquark picture.  Numerical results for the form factors are given in Sec.~IV. In addition, using the helicity amplitude method, we also carry out a phenomenological analysis of  decay widths, branching ratios and forward-backward asymmetry of the weak decays of triply heavy baryons. 
A brief summary  is given in the last section.

\section{Theoretical framework}

The theoretical framework for the baryonic transitions induced by charged current will be introduced in this section. 
Semi-leptonic and nonleptonic decays of triply heavy baryons $\Omega_{ccc}^{++}\to \Xi_{cc}^{+}/\Omega_{cc}^+$ can be calculated  with the effective Hamiltonian:
\begin{eqnarray}
	&&{\cal H}_{{\rm eff}}(c\to d/s~\ell^{+}\nu_{l})=\frac{G_{F}}{\sqrt{2}}\Big(V_{cd}^{*}[\bar{d}\gamma_{\mu}(1-\gamma_5)c][\bar{\nu}_{l}\gamma^{\mu}(1-\gamma_5)l]+V_{cs}^{*}[\bar{s}\gamma_{\mu}(1-\gamma_5)c][\bar{\nu}_{l}\gamma^{\mu}(1-\gamma_5)l]\Big),\notag\\
	&&{\cal H}_{{\rm eff}}(c\to d\bar{s}u/s\bar{d}u)=\sum_{i=1,2}\frac{G_F}{\sqrt{2}}C_i(V_{cs}V^*_{ud}O^{s\bar d u}_i+V_{cd}V^*_{us}O_i^{d\bar s u}+V_{cs}V^*_{us}O^{s\bar s u}_i+V_{cd}V^*_{ud}O_i^{d\bar d u})+h.c. ,\notag\\
	&&O^{q_1\bar q_2 q_3}_1=[\bar{q_1}_\alpha \gamma_{\mu}(1-\gamma_5)c_\beta][\bar{q_3}_\beta\gamma^{\mu}(1-\gamma_5) {q_2}_{\alpha}],\quad   O^{q_1\bar q_2 q_3}_2=[\bar{q_1}_\alpha \gamma_{\mu}(1-\gamma_5)c_\alpha][\bar{q_3}_\beta \gamma^{\mu}(1-\gamma_5){q_2}_{\beta}],\label{eq:efhvc}
\end{eqnarray}
where $G_F$ is the Fermi constant. The $C_i$ are Wilson coefficients \cite{Buchalla:1995vs} and  CKM matrix elements are taken from Ref.~\cite{Zyla:2020zbs}:
\begin{eqnarray}
	&&|V_{cd}|=0.221,\quad|V_{cs}|=0.987\quad,|V_{ud}|=0.973,\quad |V_{us}|=0.2245,\notag\\
	&&G_F=1.166\times 10^{-5}{\rm GeV}^{-2},\quad C_1(m_c)=-0.493,\quad C_2(m_c)=1.250.
\end{eqnarray}

With the help of the Hamiltonian in Eq.~\eqref{eq:efhvc},    the decay amplitude for   the semi-leptonic decay process  is given as
\begin{equation}
	i\mathcal{M}(\Omega_{ccc}^{++}\to\mathcal{B}_{cc}^{+}\;\ell^+\nu_l)=\frac{G_{F}}{\sqrt{2}}V_{cq}^{*}\langle \mathcal{B}_{cc}^{+}|\bar{q}\gamma_{\mu}(1-\gamma_5)c|\Omega_{ccc}^{++}\rangle \bar{\nu}_l\gamma^{\mu}(1-\gamma_5)l,\;\mathcal{B}_{cc}^{+}= \Xi_{cc}^{+}/\ \Omega_{cc}^+,\;q=d/s.
\end{equation}
On the one hand, 
the hadron matrix element can be parameterized by the form factors. For spin-$3/2$ to spin-$1/2$ process, the matrix elements are written as
\begin{eqnarray}
	&&	\langle{\cal B}_{cc}^{+}(P^{\prime},S^{\prime}=\frac{1}{2},S_{z}^{\prime})|\bar{q}\gamma^{\mu}(1-\gamma_{5})c|{\Omega}_{ccc}^{++}(P,S=\frac{3}{2},S_{z})\rangle \nonumber\\
	&& =  \bar{u}(P^{\prime},S_{z}^{\prime})\Big[\gamma^{\mu}P^{\prime\alpha}\frac{\mathtt{f}_{1}(q^{2})}{M}+\frac{\mathtt{f}_{2}(q^{2})}{M(M-M^\prime)}P^{\prime\alpha}P^{\mu} +\frac{\mathtt{f}_{3}(q^{2})}{M(M-M^{\prime})}P^{\prime\alpha}P^{\prime\mu}+\mathtt{f}_{4}(q^{2})g^{\alpha\mu}\Big]\gamma_{5}u_{\alpha}(P,S_{z})\nonumber\\
	& &\quad-\bar{u}(P^{\prime},S_{z}^{\prime})\Big[\gamma^{\mu}P^{\prime\alpha}\frac{\mathtt{g}_{1}(q^{2})}{M}+\frac{\mathtt{g}_{2}(q^{2})}{M(M-M^\prime)}P^{\prime\alpha}P^{\mu} +\frac{\mathtt{g}_{3}(q^{2})}{M(M-M^{\prime})}P^{\prime\alpha}P^{\prime\mu}+\mathtt{g}_{4}(q^{2})g^{\alpha\mu}\Big]u_{\alpha}(P,S_{z}).\label{eq:matrix_element_32VA} 
\end{eqnarray}
Actually, we can define another set of form factors which are called helicity form factors. A detail discussion is shown in the appendix.  On the another hand, the hadron matrix element can also be constructed by   LFQM.  

Before presenting the model calculation of form factors,   we comment on the contaminations from the spin-1/2 $\Omega_{ccc}$  state.  The $\Omega_{ccc}$ is made of three charm quarks. For the ground  state $\Omega_{ccc}$ which can only weak decay,  charm quarks are identical  fermions and thus the total wave functions are anti-symmetric under the interchange of two charm quarks. Wave functions in momentum space and flavor space are both symmetric, while it is antisymmetric in color space. The spin of the charm quarks should be symmetric, and as a result the total spin of  the lowest-lying $\Omega_{ccc}$ is 3/2. 
The orbital excited state  with $L=1$ can have spin-1/2, but its mass is naturally higher than  the ground state, and thus can either decay into the ground state through the strong or electromagnetic interaction. Such feed-down contributions will not affect the study  of decay branching fractions of the ground state, similar with the study of $B$ meson decays.

\subsection{Light-front quark model}
\begin{figure}[htp]
\includegraphics[width=0.5\columnwidth]{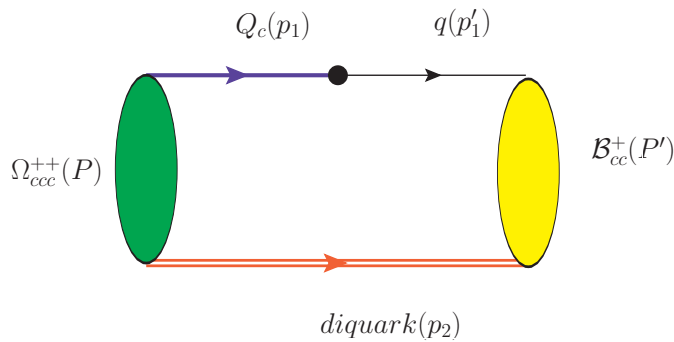} 
\caption{Feynman diagram for triply heavy baryon $\Omega_{ccc}^{++}$ decays into a  baryon $B^{+}_{cc}$ with two spectator quarks as a diquark. Here $P$ and $P^{\prime}$ are the momentum of the initial and final baryons, respectively. At the quark level, the transition is one heavy quark $Q_{c}$ with momentum $p_{1}$  into a light quark $q$ with momentum $p_{1}^{\prime}$ and the diquark with momentum $p_{2}$. The black dot denotes the weak interaction vertex.}\label{diquark}
\end{figure}

For the hadronic state which involves coordinate, color, spin and flavor distributions, one can construct the wave function in light-cone frame.  In the lightcone frame, a four-vector is represented as $v^\mu=(v^-,v^+,v_\perp)$ with $v^\pm=v^0\pm v^3$ and $v_\perp=(v^1,v^2)$. 
In our analysis the diquark picture is adopted and the Feynman diagram for the transition shown in Fig.~\ref{diquark}. 

For the $J^{P}=1/2^{+}$ baryon state, the   wave function is
\begin{eqnarray}
	|{\cal B}(P,S,S_{z})\rangle & = & \int\{d^{3}p_{1}\}\{d^{3}p_{2}\}2(2\pi)^{3}\delta^{3}(\tilde{P}-\tilde{p}_{1}-\tilde{p}_{2})\nonumber \\
	&  & \times\sum_{\lambda_{1},\lambda_{2}}\Psi^{SS_{z}}(\tilde{p}_{1},\tilde{p}_{2},\lambda_{1},\lambda_{2})|Q_c(p_{1},\lambda_{1})({\rm{di}})(p_{2},\lambda_{2})\rangle,\label{eq:state_vector}
	\end{eqnarray}
where $Q_c$ donates the charm quark and  ``$({\rm{di}})$" presents the diquark shown in Fig.~\ref{diquark}.
$\lambda_1$ and $\lambda_2$ denote their helicities,   
$P$ is the total momenta of the baryon, $p_1,~ p_2$ are the on-mass-shell light-front momentum of the heavy quark and the diquark, respectively. 
The momenta $\tilde{P}$, $\tilde{p}_{1}$ and $\tilde{p}_{2}$ are defined as
$\tilde{p}=(p^+,p_{\perp})$. Since the on shell momentum has only three degree of freedom with four component, the
minus component of the momentum can be determined from their on-shell condition as $p^{-}=(m^2+p_{\perp}^2)/p^{+}$. 
The distribution is given as
\begin{equation}
	\Psi^{SS_{z}}(\tilde{p}_{1},\tilde{p}_{2},\lambda_{1},\lambda_{2})=\frac{1}{\sqrt{2(p_{1}\cdot\bar{P}+m_{1}M_{0})}}\bar{u}(p_{1},\lambda_{1})\Gamma_{S(A)} u(\bar{P},S_{z})\phi(x,k_{\perp}),\label{eq:momentum_wave_function_1/2}
\end{equation}
and for the $J^P=3/2^+$ state the wave function is given as~\cite{Hu:2020mxk}.
\begin{equation}
	\Psi^{SS_{z}}(\tilde{p}_{1},\tilde{p}_{2},\lambda_{1},\lambda_{2})=\frac{1}{\sqrt{2(p_{1}\cdot\bar{P}+m_{1}M_{0})}}\bar{u}(p_{1},\lambda_{1})\Gamma^{\alpha}_{A}(p_{2},\lambda_{2})u_{\alpha}(\bar{P},S_{z})\phi(x,k_{\perp}),\label{eq:momentum_wave_fuction_3/2}
\end{equation}
where $\Gamma$ is the coupling vertex of the charm quark $Q_{c}$ and the diquark in the baryon state.

In the $J^P=1/2^+$ case, when the diquark is a scalar, the coupling vertex is  $\Gamma_{S}=1$.
In Ref.~\cite{Chua:2018lfa}, when an axial-vector diquark is involved, the vertex becomes
\begin{align}
	\Gamma_{A} & =\frac{\gamma_{5}}{\sqrt{3}}\left(\slashed\epsilon^{*}(p_{2},\lambda_{2})-\frac{M_0+m_1+m_2}{\bar{P}\cdot p_2+m_2M_0}\epsilon^{*}(p_{2},\lambda_{2})\cdot\bar{P}\right).\label{eq:momentum_wave_function_1/2gamma}
\end{align}
For the $J^P=3/2^+$ case, there is only axial-vector diquark involved. The spin of diquark must be $1$  since the total spin is coupled to $3/2$. The  coupling vertex $\Gamma^{\alpha}_{A}$ is 
\begin{equation}
	\Gamma_{A}^{\alpha}=-\left(\epsilon^{*\alpha}(p_{2},\lambda_{2})-\frac{p_2^{\alpha}}{\bar{P}\cdot p_2+m_2M_0}\epsilon^{*}(p_{2},\lambda_{2})\cdot\bar{P}\right).
\end{equation}
Here $ m_1$ and $m_2$ are the masses of the heavy quark and the diquark.
The $\bar P$ is the sum of the on-mass-shell momenta of the charm quark $Q_c$ and diquark. Thus we have $\bar{P}=p_1+p_2$ and $\bar{P}^2=M_{0}^2$. 
The  invariant mass $M_0$ will be different from the baryon mass $M$, since the quark ,diquark and the baryon they composed can not be on their mass shells simultaneously.
The momentum $P$ and mass $M$ of baryon will satisfy $M^2=P^2$ which is the physical mass-shell condition.
But the momentum $\bar P$ is not equal to the $P$.

The $\phi$ is a Gaussian-type function in wave function $\Psi^{SS_{z}}(\tilde{p}_{1},\tilde{p}_{2},\lambda_{1},\lambda_{2})$:
\begin{equation}
	\phi=4\left(\frac{\pi}{\beta^{2}}\right)^{3/4}\sqrt{\frac{e_{1}e_{2}}{x_{1}x_{2}M_{0}}}\exp\left(\frac{-\vec{k}^{2}}{2\beta^{2}}\right),\label{eq:Gauss}
\end{equation}
where $e_1$ and $e_2$ represent the energy of the charm quark $Q_c$ and diquark in the rest frame of $\bar{P}$.
$x_1$ and $x_2$ are the light-front momentum fractions which is satisfying $0<x_2<1$ and $x_1+x_2=1$.
For brevity we introduce the internal momenta to describe the internal motion of the constituent quarks as
\begin{eqnarray}
	 &&k_i=(k_i^-,k_i^+,k_{i\bot})=(e_i-k_{iz},e_i+k_{iz},k_{i\bot})=
  (\frac{m_i^2+k_{i\bot}^2}{x_iM_0},x_iM_0,k_{i\bot}),\nonumber\\
	&&p^+_1=x_1\bar P^+, \qquad\qquad p^+_2=x_2 \bar P^+,
 \qquad\qquad \nonumber\\
	&&p_{1\perp}=x_1 \bar P_{\perp}+k_{1\perp},
  ~~~ p_{2\perp}=x_2 \bar P_{\perp}+k_{2\perp},
  ~~~ k_{\perp}=-k_{1\perp}=k_{2\perp}.
\end{eqnarray}
The $\vec k$ in the Eq.~\eqref{eq:Gauss} is the internal three-momentum vector of diquark, and $\vec k=(k_{2\bot},k_{2z})=(k_{\bot}, k_{z})$.
The parameter $\beta$ in Eq.~(\ref{eq:Gauss})  describes the momentum distributions among the constituent quarks and is $\beta_{c[cc]}=0.553,\beta_{d[cc]}=0.370,\beta_{s[cc]}=0.435$.  Since the previous  predictions  in LFQM for doubly heavy baryon decays in Ref.~\cite{Hu:2020mxk} are typically larger than results from other method such as QCD sum rules~\cite{Shi:2019hbf}, we have tuned down the shape parameters  $\beta$s compared to the previous work~\cite{Hu:2020mxk}. 
Using the internal momentum we can express the invariant mass square $M_0^2$ as a function of the internal variables $(x_{i},k_{i\bot})$,
 \begin{eqnarray} \label{eq:Mpz}
	  M_0^2=\frac{k_{1\perp}^2+m_1^2}{x_1}+ \frac{k_{2\perp}^2+m_2^2}{x_2}.
 \end{eqnarray}
Then the expressions of the energy  $e_i$ and $k_z$ in terms of the internal variables $(x_{i},k_{i\bot})$ are
\begin{eqnarray}
 	e_i&=&\frac{x_iM_0}{2}+\frac{m_i^2+k_{i\perp}^2}{2x_iM_0} =\sqrt{m_i^2+k_{i\bot}^2+k_{iz}^2},\qquad k_{iz}=\frac{x_iM_0}{2}-\frac{m_i^2+k_{i\perp}^2}{2x_iM_0}.
 \end{eqnarray}
In the following,  the notation $x=x_2$ and  $x_1=1-x$ will be adopted.

With the help of the expression of  hadron state, the matrix element can be calculated and the form factors can be solved.
The hadron matrix element is
\begin{eqnarray}
	&  & \langle{\cal B}^{+}_{cc}(P^{\prime},S^{\prime}=\frac{1}{2},S_{z}^{\prime})|\bar{q}\gamma^{\mu}(1-\gamma_{5})Q_{c}|{\Omega}_{ccc}^{++}(P,S=\frac{3}{2},S_{z})\rangle\nonumber \\
	& = & \int\{d^{3}p_{2}\}\frac{\phi^{\prime}(x^{\prime},k_{\perp}^{\prime})\phi(x,k_{\perp})}{2\sqrt{p_{1}^{+}p_{1}^{\prime+}(p_{1}\cdot\bar{P}+m_{1}M_{0})(p_{1}^{\prime}\cdot\bar{P}^{\prime}+m_{1}^\prime M_{0}^{\prime})}}\nonumber \\
	&  & \times\sum_{\lambda_{2}}\bar{u}(\bar{P}^{\prime},S_{z}^{\prime})\left[\bar{\Gamma}^{\prime}_{A}(\slashed p_{1}^{\prime}+m_{1}^{\prime})\gamma^{\mu}(1-\gamma_{5})(\slashed p_{1}+m_{1})\Gamma^{\alpha}_{A}\right]u_\alpha(\bar{P},S_{z}),\label{eq:matrix_element_onehalf}
\end{eqnarray}
where $q=d,s$ and
\begin{eqnarray}
	&&m_{1}=m_{c},\quad m_{1}^{\prime}=m_{q},\quad m_{2}=m_{(di)},\nonumber\\
	&&\bar{P}=p_{1}+p_{2},\quad\bar{P}=p_{1}^{\prime}+p_{2},\quad M_{0}^2=\bar{P}^{2},\quad M_{0}^{\prime2}=\bar{P}^{\prime2}.
\end{eqnarray}
The $p_{1}$ and $p_{1}^{\prime}$ are the four-momentum of the charm quark and final quark. 
The $p_2$ is the momentum of spectator diquark which is shown in Fig.~\ref{diquark}. 
$P$ and $P^{\prime}$ are the four-momentum of the initial baryons ${\Omega}_{ccc}^{++}$ and final baryon states ${\cal B}^{+}_{cc}$, respectively. 
$M$ and $M^{\prime}$ are the physical masses of them. It is noted that  $M_{0}$ and $M^{\prime}_{0}$ are the invariant masses of the initial and final baryon states with quark and diquark component which are different from $M$ and $M^{\prime}$. The $\bar\Gamma$ is defined as
\begin{eqnarray}
	\Gamma_{S}&=&\bar{\Gamma}^{\prime}_{S}=1,\notag\\
	\bar{\Gamma}^{\prime}_{A}  &=&\frac{1}{\sqrt{3}}\left(-\slashed\epsilon(p_{2},\lambda_{2})+\frac{M_0^{\prime}+m_1^{\prime}+m_2}{\bar{P}^{\prime}\cdot p_2+m_2M_0^{\prime}}\epsilon(p_{2},\lambda_{2})\cdot\bar{P}^{\prime}\right)\gamma_{5}.\label{axial-vector diquarkprime}
\end{eqnarray}

To extract  the eight form factors,  one can employ eight different structures $\bar{u}_{\beta}(P,S_{z})(\Gamma ^{\mu\beta})_{i}/(\Gamma_{5}^{\mu\beta})_{i}u(P^{\prime},S_{z}^{\prime})$  with $(\Gamma^{\mu\beta})_{i}=\{\gamma^{\mu}P^{\prime\beta},P^{\prime\mu}P^{\prime\beta},P^{\mu}P^{\prime\beta},g^{\mu\beta}\}$ and $(\Gamma_5^{\mu\beta})_{i}=\{\gamma^{\mu}P^{\prime\beta}\gamma_{5},P^{\prime\mu}P^{\prime\beta}\gamma_{5},P^{\mu}P^{\prime\beta}\gamma_{5},g^{\mu\beta}\gamma_5\}$. Multiplying the above structures on both sides of Eq.~\eqref{eq:matrix_element_onehalf} and summing the polarizations lead to  eight equations. For the vector current form factors,   four equations can be obtained by multiply the vertex $(\Gamma_5^{\mu\beta})_{i}$: 
\begin{eqnarray}
	&&{\rm Tr}\Big\{ (\slashed P^\prime+M)\Big[\gamma^{\mu}P^{\prime\alpha}\frac{\mathtt{f}_{1}(q^{2})}{M}+\frac{\mathtt{f}_{2}(q^{2})}{M(M-M^\prime)}P^{\prime\alpha}P^{\mu}+\frac{\mathtt{f}_{3}(q^{2})}{M(M-M^{\prime})}P^{\prime\alpha}P^{\prime\mu}
+\mathtt{f}_{4}(q^{2})g^{\alpha\mu}\Big]\gamma_{5} u_{\alpha}(P,S_{z})\bar{u}_{\beta}(P,S_{z})(\Gamma_{5}^{\mu\beta})_{i}\Big\}\nonumber\\
 	&&= \int\{d^{3}p_{2}\}\frac{\phi^{\prime}(x^{\prime},k_{\perp}^{\prime})\phi(x,k_{\perp})}{2\sqrt{p_{1}^{+}p_{1}^{\prime+}(p_{1}\cdot\bar{P}+m_{1}M_{0})(p_{1}^{\prime}\cdot\bar{P}^{\prime}+m_{1}^{\prime}M_{0}^{\prime})}}\nonumber \\
	&  &\quad \times\sum_{S_{z}^{\prime}\lambda_{2}}{\rm Tr}\Big\{ (\slashed P^\prime+M)\bar{\Gamma}^{\prime}_{A}(\slashed p_{1}^{\prime}+m_{1}^{\prime})\gamma_{\mu}(\slashed p_{1}+m_{1})\Gamma^\alpha_{A}u_{\alpha}(P,S_{z})\bar{u}_{\beta}(P,S_{z})(\bar{\Gamma}_{5}^{\mu\beta})_{i}\Big\},\label{eq:Fisolving} 
\end{eqnarray}
and the equations for the axis vectors form factors can be constructed in a similar  way.
Using the formula in Ref.~\cite{Chua:2019yqh},  one can give the spin sum of the $3/2$ spinors as
\begin{eqnarray}
	\sum_{S_{z}=-3/2}^{3/2}u_{\alpha}(\bar{P},S_{z})\bar{u}_{\beta}(\bar{P},S_{z})=-(\bar{\slashed P}+M_{0})\Big[G_{\beta\alpha}(\bar{P})-\frac{1}{3}G_{\alpha\sigma}(\bar{P})G_{\beta\lambda}(\bar{P})\gamma^{\sigma}\gamma^{\lambda}\Big],\nonumber
\end{eqnarray} with
\begin{eqnarray}
	G_{\beta\alpha}(\bar{P})=g_{\beta\alpha}-\frac{\bar{P}_{\beta}\bar{P}_{\alpha}}{M_{0}^{2}}.\nonumber
\end{eqnarray}

Solving these equations, one can express these form factors by LFQM as
\begin{eqnarray}
	{\mathtt{f}_{1}}(q^2)&=&\frac{-M^2}{2s_-^2s_+^2}\big\{4{M}\big[{H_1}s_--{H_2}{M}\big]-2{H_3}\big(s_--2MM^\prime\big)+{H_4}(s_-s_+)\big\},\notag\\
	{\mathtt{f}_{2}}(q^2)&=&\frac{-{M}\bar M}{s_-^3s_+^2}\big\{s_-\big[H_1M\big(s_--2MM^\prime\big)+H_4\big(s_-+MM^\prime\big)s_+\big]+4H_2M^2\big(2s_+-3MM^\prime\big)\notag\\
	&&-2H_3\big[M^4-2M^3M^\prime+2M^2(6M^{\prime 2}-q^2)+2M M^\prime(q^2-M^{\prime 2})+(M^{\prime2}-q^2)^2\big]\big\},\notag\\
	{\mathtt{f}_{3}}(q^2)&=&\frac{{M^3}{\bar M}}{s_-^3s_+^2}\big\{s_-\big(H_4s_+-2H_1 M\big)+20H_2M^2-4H_3\big(2s_+-3MM^\prime\big)\big\},\notag\\
	{\mathtt{f}_{4}}(q^2)&=&\frac{1}{2s_-^2s_+}\big\{ s_-\left[{H_1}M+{H_4}s_+\right]+2{H_2}{M}^2-2{H_3}\big[s_-+MM^\prime\big]\big\},\label{ffe}
\end{eqnarray}
where $\bar M=M-M^\prime$ and $H_i$ is defined as follows
\begin{eqnarray}
	H_{i} & = &  \int\{d^{3}p_{2}\}\frac{\phi^{\prime}(x^{\prime},k_{\perp}^{\prime})\phi(x,k_{\perp})}{2\sqrt{p_{1}^{+}p_{1}^{\prime+}(p_{1}\cdot\bar{P}+m_{1}M_{0})(p_{1}^{\prime}\cdot\bar{P}^{\prime}+m_{1}^{\prime}M_{0}^{\prime})}}\nonumber \\
	&  &\quad \times\sum_{S_{z}^{\prime}\lambda_{2}}{\rm Tr}\Big\{ (\slashed P^\prime+M)\bar{\Gamma}^{\prime}_{A}(\slashed p_{1}^{\prime}+m_{1}^{\prime})\gamma_{\mu}(\slashed p_{1}+m_{1})\Gamma^\alpha_{A}u_{\alpha}(P,S_{z})\bar{u}_{\beta}(P,S_{z})(\bar{\Gamma}_{5}^{\mu\beta})_{i}\Big\}.\label{eq:Fi_L}
\end{eqnarray}

One can obtain the form factors $\mathtt{g}_{1,2,3,4}$ by the same method, and four equations can be constructed by multiply $(\Gamma^{\mu\beta})_{i}$. The form factors $\mathtt{g}_{1,2,3,4}$ are expressed as
\begin{eqnarray}
	{\mathtt{g}_{1}}(q^2)&=&\frac{-M^2}{2s_-^2s_+^2}\big\{4{M}\big[{K_1}s_+-{K_2}{M}\big]-2{K_3}\big(s_++2MM^\prime\big)+{K_4}(s_-s_+)\big\},\notag\\
	{\mathtt{g}_{2}}(q^2)&=&\frac{{M}\bar M}{s_+^3s_-^2}\big\{s_+\big[K_1M\big(s_++2MM^\prime\big)+K_4\big(s_+-MM^\prime\big)s_+\big]+4K_2M^2\big(2s_-+3MM^\prime\big)\notag\\
	&&-2K_3\big[M^4+2M^3M^\prime+2M^2(6M^{\prime 2}-q^2)-2M M^\prime(q^2-M^{\prime 2})+(M^{\prime2}-q^2)^2\big]\big\},\notag\\
	{\mathtt{g}_{3}}(q^2)&=&\frac{-{M^3}{\bar M}}{s_+^3s_-^2}\big\{s_+\big(K_4s_--2K_1 M\big)+20K_2M^2-4K_3\big(2s_-+3MM^\prime\big)\big\},\notag\\
	{\mathtt{g}_{4}}(q^2)&=&\frac{1}{2s_+^2s_-}\big\{ s_+\left[{K_1}M+{K_4}s_-\right]+2{K_2}{M}^2-2{K_3}\big[s_+-MM^\prime\big]\big\},\label{ffe}
\end{eqnarray}
The expression of $\mathtt{g}_{i}$ can also be obtained  by applying the transformation to $\mathtt{f}_{i}$ as
\begin{eqnarray}
	&&s_\pm\to s_\mp,\quad M^{\prime}\to -M^{\prime},\quad \bar M\to \bar M ,\quad H_{i}\to K_{i}, \qquad i=1,\notag\\
	&&s_\pm\to s_\mp,\quad M^{\prime}\to -M^{\prime},\quad \bar M\to \bar M ,\quad H_{i}\to -K_{i}, \qquad i=2,3,4,
\end{eqnarray} 
where $K_i$ is defined as
\begin{eqnarray}
	K_{i} & = & \int\{d^{3}p_{2}\}\frac{\phi^{\prime}(x^{\prime},k_{\perp}^{\prime})\phi(x,k_{\perp})}{2\sqrt{p_{1}^{+}p_{1}^{\prime+}(p_{1}\cdot\bar{P}+m_{1}M_{0})(p_{1}^{\prime}\cdot\bar{P}^{\prime}+m_{1}^{\prime}M_{0}^{\prime})}}\nonumber \\
	&  &\quad \times\sum_{S_{z}^{\prime}\lambda_{2}}{\rm Tr}\Big\{ (\slashed P^\prime+M)\bar{\Gamma}^{\prime}_{A}(\slashed p_{1}^{\prime}+m_{1}^{\prime})\gamma_{\mu}\gamma_5(\slashed p_{1}+m_{1})\Gamma^\alpha_{A}u_{\alpha}(P,S_{z})\bar{u}_{\beta}(P,S_{z})(\bar{\Gamma}_{5}^{\mu\beta})_{i}\Big\}.\label{eq:Gi_1}
\end{eqnarray}

Generally, the amplitude of spin- $3/2$ to spin-$1/2$ process can be derived by applying the time reversal transformation to the amplitude of spin-$1/2$ to spin-$3/2$ processes.
Since the definition of form factor is similar to the time reversal form factor of spin-$1/2$ to spin-$3/2$ processes in the previous works~\cite{Hu:2020mxk}, the expression in Eq.~\eqref{ffe} can be derived by applying a transformation to previous results as
\begin{eqnarray}
	&&(\mathtt{f}^{\frac{3}{2}\to\frac{1}{2}}_{1},\mathtt{f}^{\frac{3}{2}\to\frac{1}{2}}_{2},\mathtt{f}^{\frac{3}{2}\to\frac{1}{2}}_{3},\mathtt{f}^{\frac{3}{2}\to\frac{1}{2}}_{4})=T_f\bigg[(\mathtt{f}^{\frac{1}{2}\to\frac{3}{2}}_{1}\frac{M}{M^\prime},-\mathtt{f}^{\frac{1}{2}\to\frac{3}{2}}_{3}\frac{M^\prime}{\bar M} ,-\mathtt{f}^{\frac{1}{2}\to\frac{3}{2}}_{2}\frac{M}{\bar M} ,-\mathtt{f}^{\frac{1}{2}\to\frac{3}{2}}_{4} )\bigg],\notag\\
	&&(\mathtt{g}^{\frac{3}{2}\to\frac{1}{2}}_{1},\mathtt{g}^{\frac{3}{2}\to\frac{1}{2}}_{2},\mathtt{g}^{\frac{3}{2}\to\frac{1}{2}}_{3},\mathtt{g}^{\frac{3}{2}\to\frac{1}{2}}_{4})=T_g\bigg[(\mathtt{g}^{\frac{1}{2}\to\frac{3}{2}}_{1}\frac{M}{M^\prime},\mathtt{g}^{\frac{1}{2}\to\frac{3}{2}}_{3}\frac{M^\prime}{\bar M} ,\mathtt{g}^{\frac{1}{2}\to\frac{3}{2}}_{2}\frac{M}{\bar M} ,\mathtt{g}^{\frac{1}{2}\to\frac{3}{2}}_{4} \frac{M}{M^\prime})\bigg],\notag\\
	&&T_f: M\leftrightarrow M^\prime,\;(H_{1},H_{2},H_3,H_4)\to(H_{1},-H_{3},-H_2,-H_4),\quad T_g: M\leftrightarrow M^\prime.\label{trr}
\end{eqnarray}

Using the momenta space wave function, one can express the momentum distribution of quarks in hadron. 
However the physical states also contain the color, flavor and spin informations. 
As we mentioned above, when we consider the leading order contribution which without  the gluon exchange, the color space is trivial. 
Therefore we will only discuss the flavor-spin wave function below.
Since  charm quarks in $\Omega_{ccc}^{++}$ are identical, the flavor wave function $ccc$ is complete symmetry under the interchange of the three constituent quarks. Thus the spin wave function can only be the spin-3/2 states since it is only complete symmetry state.  Using the diquark basis $[c_1,c_2]_A=(c_1c_2+c_2c_1)/\sqrt{2}$, one can express the flavor-spin wave function of baryons for example as
    \begin{eqnarray}
|{\cal B}_{cc}(S=\frac{1}{2}, S_z=+1/2)\rangle&=&\frac{1}{\sqrt{2}}(c_1c_2+c_2c_1)q\times\bigg(\frac{1}{\sqrt{6}}(\uparrow\downarrow\uparrow+\downarrow\uparrow\downarrow-2\uparrow\uparrow\downarrow)\bigg)=-[c_1c_2]_A\times q,\quad q=d,s,\notag\\
|\Omega_{ccc}^{++}(S=\frac{3}{2}, S_z=+3/2)\rangle&=&\frac{1}{\sqrt 6}(c_1c_2c_3+c_1c_3c_2+c_2c_1c_3+c_2c_3c_1+c_3c_2c_1+c_3c_1c_2)\times (\uparrow\uparrow\uparrow)\notag\\
&=&\frac{1}{\sqrt 3}([c_1c_2]_Ac_3+[c_1c_3]_Ac_2+[c_2c_3]_Ac_1).
  \end{eqnarray}
Thus estimating the overlap of the flavor-spin wave function $\langle{\cal B}_{cc}|\Omega_{ccc}^{++}\rangle$, a factor $-\sqrt{3}$ should be included.

\section{Numerical results}

\subsection{Form factors}
With the help of momenta space and flavor-spin wave function in LFQM, the form factors defined in Eq.~\eqref{eq:matrix_element_32VA} can be estimated. 
In our calculation, the masses of quark are taken from Refs.~\cite{Li:2010bb,Verma:2011yw,Shi:2016gqt},
\begin{eqnarray}
	 m_u=m_d= 0.25~{\rm GeV}, \;\; m_s=0.37~{\rm GeV}, \;\; m_c=1.4~{\rm GeV}.\label{eq:mass_quark}
\end{eqnarray}
The mass of diquark is approximatively taken as,
\begin{eqnarray}
	m_{[cc]}=2m_{c}.\label{eq:mass_diquark}
\end{eqnarray}
The masses of the initial and final hadron states are used as \cite{Faustov:2021qqf,Hu:2020mxk,Zyla:2020zbs}
\begin{eqnarray}
	&&m_{\Omega_{ccc}^{++}}=4.712\rm{GeV},\quad m_{\Xi_{cc}^{+}}=3.621\rm{GeV},\quad m_{\Omega_{cc}^{+}}=3.738\rm{GeV},\notag\\
	&&m_\pi=0.130\rm{GeV},\quad m_K=0.494\rm{GeV}.
\end{eqnarray}
In Ref.~\cite{Wang:2018utj}  the lifetime of $\Omega_{ccc}^{++}$ has been estimated with the next-to-leading order QCD corrections taken into account:
\begin{eqnarray} \tau_{\Omega_{ccc}^{++}}=164\times 10^{-15}s.
\end{eqnarray}

For these processes induced by charm quark decays, the $q^2$ dependence of form factors are represented by single pole model \cite{Zhao:2018mrg}
\begin{align}
	&F(q^{2})=\frac{F(0)}{1-\frac{q^{2}}{m_{{\rm pole}}^{2}}},\label{eq:main_fit_formula_cdecay}
\end{align}
where ${m_{\rm pole}} = m_D=1.869\rm{GeV}$ for $c\to d$ process and ${m_{\rm pole}} = m_{D_s}=1.968\rm{GeV}$ for $c\to s$ process \cite{Zyla:2020zbs}.  
Numerical results of processes $\Omega_{ccc}^{++}\to\Xi_{cc}^+/\Omega_{cc}^+\ell^+\nu_l$ are collected in Table.~\ref{ff}. The error of these form factors come from parameter $\beta_{c[cc]},\;\beta_{q[cc]}$ and $m_{(di)}$ respectively. 
These parameter are varied by $10\%$.

\begin{table}[!htb]
\caption{Numerical results for form factors at $q^2=0$ in $\Omega_{ccc}^{++}$ decays.}
\label{ff} %
\begin{tabular}{|c|c|c|c|c|c|c|c|c|}
\hline \hline
\multicolumn{4}{|c|}{$\Omega_{ccc}^{++}\to\Omega_{cc}^+$}\tabularnewline\hline
$\mathtt{f}_{1}(0)$&$-0.11\pm0.04\pm0.01\pm0.04$&$\mathtt{g}_{1}(0)$&$-17.00\pm1.58\pm 4.48\pm0.57$\tabularnewline\hline
$\mathtt{f}_{2}(0)$&$-11.43\pm0.29\pm0.36\pm1.97$&$\mathtt{g}_{2}(0)$&$0.05\pm0.02\pm0.03\pm0.02$\tabularnewline\hline
$\mathtt{f}_{3}(0)$&$28.36\pm0.74\pm2.38\pm3.27$&$\mathtt{g}_{3}(0)$&$4.02\pm0.11\pm0.36\pm0.32$\tabularnewline\hline
$\mathtt{f}_{4}(0)$&$4.10\pm0.14\pm0.30\pm0.01$&$\mathtt{g}_{4}(0)$&$0.81\pm0.05\pm0.03\pm0.09$\tabularnewline\hline\hline
\multicolumn{4}{|c|}{$\Omega_{ccc}^{++}\to\Xi_{cc}^+$}\tabularnewline\hline
$\mathtt{f}_{1}(0)$&$-0.31\pm0.12\pm0.064\pm0.06$&$\mathtt{g}_{1}(0)$&$1.74\pm0.26\pm0.23\pm0.14$\tabularnewline\hline
$\mathtt{f}_{2}(0)$&$-11.64\pm0.20\pm0.89\pm0.91$&$\mathtt{g}_{2}(0)$&$-2.73\pm0.20\pm0.09\pm0.23$\tabularnewline\hline
$\mathtt{f}_{3}(0)$&$21.45\pm0.32\pm1.82\pm0.95$&$\mathtt{g}_{3}(0)$&$2.67\pm0.17\pm0.51\pm0.69$\tabularnewline\hline
$\mathtt{f}_{4}(0)$&$3.84\pm0.10\pm0.30\pm0.03$&$\mathtt{g}_{4}(0)$&$0.82\pm0.03\pm 0.05\pm 0.06$\tabularnewline\hline
\end{tabular}
\end{table}

We can see that the form factors $\mathtt{g}_{1,2}$ in two channels have an obvious difference. We have checked that the difference mainly comes from the $K_1$ which in influenced by the mass of final baryon states. It can be seen that the $K_1$ give a large contribution in expression of $\mathtt{g}_{1,2}$ in Eq.~\eqref{ffe}, because the coefficient of $K_1$ are  proportional to $s_+ M$.  Thus the difference of final baryon states mass is the main source of SU(3) symmetry breaking effects. 
\subsection{Semileptonic decays}

In this subsection, we use the helicity amplitude to calculate the hadron matrix element. 
The definition of helicity amplitude are
 \begin{equation}
	HV_{\lambda^{\prime},\lambda_{W}}^{\lambda}\equiv\langle{\cal B}_{cc}^{+}(\lambda^{\prime})|\bar{q}\gamma^{\mu}Q_c|{\Omega}_{ccc}^{++}(\lambda)\rangle\epsilon_{W\mu}^{*}(\lambda_{W})\quad
{\rm and}\quad HA_{\lambda^{\prime},\lambda_{W}}^{\lambda}\equiv\langle{\cal B}_{cc}^{+}(\lambda^{\prime})|\bar{q}\gamma^{\mu}\gamma_{5}Q_c|{\Omega}_{ccc}^{++}(\lambda)\rangle\epsilon_{W\mu}^{*}(\lambda_{W}),
\end{equation}
where $\epsilon_{\mu}$ is the polarization vector of the virtual propagator and $\lambda_{W}$ means the polarization of the virtual propagator. 
$\lambda$ and $\lambda^{\prime}$ are the helicities of the initial and final baryon states, respectively.

Using the definition, one can give the expression of hadron helicity amplitude as
  \begin{eqnarray}
	&&HV_{\frac{1}{2},0}^{\frac{1}{2}}=\frac{\sqrt{s_-}}{2\sqrt{6 q^2}M^2}\big[2s_+(M^\prime-M)\mathtt{f}_{1}+\frac{s_-s_+}{M-M^\prime}(\mathtt{f}_{2}+\mathtt{f}_{3})-2M(M^2-M^{\prime2}+q^2)\mathtt{f}_{4}\big],\notag\\
	&&HV_{\frac{1}{2},t}^{\frac{1}{2}}=\frac{\sqrt{s_+}s_-}{2\sqrt{6 q^2}M^2}\big[-2(M^\prime+M)\mathtt{f}_{1}+(M+M^{\prime})(\mathtt{f}_{2}+\mathtt{f}_{3})+\frac{q^2}{M-M^\prime}(\mathtt{f}_{2}-\mathtt{f}_{3})-2M\mathtt{f}_{4}\big],\notag\\
	&&HV_{-\frac{1}{2},-1}^{\frac{1}{2}}=\sqrt{\frac{s_-}{3}}\big[\mathtt{f}_{1}\frac{s_+}{M^2}+\mathtt{f}_{4}\big],\quad HV_{\frac{1}{2},-1}^{\frac{3}{2}}=-\sqrt{s_-}\mathtt{f}_{4},\quad HV_{-\lambda^{\prime},-\lambda_{W}}^{-\lambda}=-HV_{\lambda^{\prime},\lambda_{W}}^{\lambda},
\end{eqnarray}
 and 
  \begin{eqnarray}
	&&HA_{\frac{1}{2},0}^{\frac{1}{2}}=-\frac{\sqrt{s_+}}{2\sqrt{6 q^2}M^2}\big[2s_-(M^\prime+M)\mathtt{g}_{1}+\frac{s_-s_+}{M-M^\prime}(\mathtt{g}_{2}+\mathtt{g}_{3})-2M(M^2-M^{\prime2}+q^2)\mathtt{g}_{4}\big],\notag\\
	&&HA_{\frac{1}{2},t}^{\frac{1}{2}}=-\frac{\sqrt{s_-}s_+}{2\sqrt{6 q^2}M^2}\big[-2(M^\prime-M)\mathtt{g}_{1}+(M+M^{\prime})(\mathtt{g}_{2}+\mathtt{g}_{3})+\frac{q^2}{M-M^\prime}(\mathtt{g}_{2}-\mathtt{g}_{3})-2M\mathtt{g}_{4}\big],\notag\\
	&&HA_{-\frac{1}{2},-1}^{\frac{1}{2}}=-\sqrt{\frac{s_+}{3}}\big[\mathtt{g}_{1}\frac{s_-}{M^2}-\mathtt{g}_{4}\big],\quad HA_{\frac{1}{2},-1}^{\frac{3}{2}}=\sqrt{s_+}\mathtt{g}_{4},\quad HA_{-\lambda^{\prime},-\lambda_{W}}^{-\lambda}=HA_{\lambda^{\prime},\lambda_{W}}^{\lambda}.\label{ha}
\end{eqnarray}
Here $s_\pm=(M\pm M^\prime)^2-q^2$.
For the semi-leptonic decays the differential decay width can be written as
\begin{eqnarray}
	\frac{d^2\Gamma}{dq^2d\cos\theta}=\frac{G_F^2|V_{CKM}^2|}{2 }\frac{\sqrt{s_-s_+}q^2}{512\pi^3M^3}\bigg[L_1+L_2\cos\theta+L_3\cos2\theta\bigg],\label{angle}
\end{eqnarray}
where $\theta$ is defined as the angle between the final leptonic $\ell^+$ and z-axis in the rest frame of $\ell^+\nu_l$ and the coefficients $L_i$ are
\begin{eqnarray}
	L_1&=&\frac{q^2}{4}(1-\hat{m_\ell}^2) \bigg((3+\hat{m_\ell}^2)(|H_{\frac{1}{2},-1}^{\frac{3}{2}}|^2+|H_{-\frac{1}{2},1}^{-\frac{3}{2}}|^2+|H_{-\frac{1}{2},-1}^{\frac{1}{2}}|^2+|H_{\frac{1}{2},1}^{-\frac{1}{2}}|^2)+2(1+\hat{m_\ell}^2)(|H_{\frac{1}{2},0}^{\frac{1}{2}}|^2+|H_{-\frac{1}{2},0}^{-\frac{1}{2}}|^2)\notag\\
	&&+4\hat{m_\ell}^2 (|H_{\frac{1}{2},t}^{\frac{1}{2}}|^2+|H_{-\frac{1}{2},t}^{-\frac{1}{2}}|^2)\bigg),\notag\\
	L_2&=&q^2(1-\hat{m_\ell}^2)\bigg(\big(|H_{\frac{1}{2},-1}^{\frac{3}{2}}|^2-|H_{-\frac{1}{2},1}^{-\frac{3}{2}}|^2+|H_{-\frac{1}{2},-1}^{\frac{1}{2}}|^2-|H_{\frac{1}{2},1}^{-\frac{1}{2}}|^2\big)+2\hat{m_\ell}^2\big(\mathcal{R}_e(H_{\frac{1}{2},t}^{\frac{1}{2}}H_{\frac{1}{2},0}^{*\frac{1}{2}} )+\mathcal{R}_e(H_{-\frac{1}{2},t}^{-\frac{1}{2}}H_{-\frac{1}{2},0}^{*-\frac{1}{2}} )\big)\bigg),\notag\\
		L_3&=&\frac{q^2}{4}(1-\hat{m_\ell}^2)^2\bigg((|H_{\frac{1}{2},-1}^{\frac{3}{2}}|^2+|H_{-\frac{1}{2},1}^{-\frac{3}{2}}|^2+|H_{-\frac{1}{2},-1}^{\frac{1}{2}}|^2+|H_{\frac{1}{2},1}^{-\frac{1}{2}}|^2)-2(|H_{\frac{1}{2},0}^{\frac{1}{2}}|^2+|H_{-\frac{1}{2},0}^{-\frac{1}{2}}|^2)\bigg).
\end{eqnarray}
Here $H_{\lambda^{\prime},\lambda_{W}}^{\lambda}=HV_{\lambda^{\prime},\lambda_{W}}^{\lambda}-HA_{\lambda^{\prime},\lambda_{W}}^{\lambda}$ and $\hat{m_\ell}^2=m^2_\ell/q^2$.
After integrated the angle $\theta$, the $q^2$ distribution differential decay width is
\begin{eqnarray}
	\frac{d\Gamma_L}{dq^2}&=&\frac{G_F^2|V_{CKM}^2|}{2 }\frac{\sqrt{s_-s_+}q^2(1-\hat{m_\ell}^2)}{768\pi^3M^3}\bigg((\hat{m_{\ell}}^2+2)(|H_{\frac{1}{2},0}^{\frac{1}{2}}|^2+|H_{-\frac{1}{2},0}^{-\frac{1}{2}}|^2)+3\hat{m_\ell}^2(|H_{\frac{1}{2},t}^{\frac{1}{2}}|^2+|H_{-\frac{1}{2},t}^{-\frac{1}{2}}|^2)\bigg),\notag\\
	\frac{d\Gamma_T}{dq^2}&=&\frac{G_F^2|V_{CKM}^2|}{2 }\frac{\sqrt{s_-s_+}q^2(1-\hat{m_\ell}^2)}{768\pi^3M^3}(1-\hat{m_\ell}^2)(\hat{m_\ell}^2+2)\bigg(|H_{\frac{1}{2},-1}^{\frac{3}{2}}|^2+|H_{-\frac{1}{2},1}^{-\frac{3}{2}}|^2+|H_{-\frac{1}{2},-1}^{\frac{1}{2}}|^2+|H_{\frac{1}{2},1}^{-\frac{1}{2}}|^2\bigg),\notag\\
		\frac{d\Gamma}{dq^2}&=&\frac{G_F^2|V_{CKM}^2|}{2 }\frac{\sqrt{s_-s_+}(1-\hat{m_\ell}^2)}{512\pi^3M^3}(2L_{1}-\frac{3}{2}L_{3}).\label{gamma}
\end{eqnarray}
One can give the total decay width by integrating the $q^2$ and  the branching ratio by using the lifetime of $\Omega_{ccc}^{++}$ in Table.~\ref{semilepton}. The first three uncertainties  in branching fractions and decay widths come from those in form factors and the last error comes from varying the  $m_{\rm pole}$ by $5\%$. It can be seen that our prediction for  $\Omega_{ccc}^{++}\to\Omega_{cc}^+e^+\nu_e$ is similar to the doubly heavy baryon decays $\Gamma(\Xi_{cc}^{++}\to \Xi_{c}^+ \ell\nu_\ell)=8.74\times 10^{-14}\rm{GeV}$~\cite{Hu:2020mxk}, since they are both induced by $c\to s$ process. 

\begin{table}[!htb]
\caption{Numerical results of decay width and branching fraction in $\Omega_{ccc}^{++}$ semi-leptonic decays.}
\label{semilepton} %
\begin{tabular}{|c|c|c|c|c|c|c|c|c|}
\hline \hline
channel & $\Gamma(\times10^{-14}\rm{GeV})$   & $\Gamma_L/\Gamma_T$& Branching fraction ($\%$)  \cite{Zhao:2022vfr} \tabularnewline
\hline
$\Omega_{ccc}^{++}\to\Xi_{cc}^+e^+\nu_e$&$0.32\pm0.04\pm0.04\pm0.05\pm0.01$&0.64&$0.08\pm0.01\pm0.01\pm0.01\pm0.00$
\tabularnewline
\hline
$\Omega_{ccc}^{++}\to\Xi_{cc}^+\mu^+\nu_\mu$&$0.31\pm0.03\pm0.04\pm0.05\pm0.01$&0.61&$0.08\pm0.01\pm0.01\pm0.01\pm0.00$\tabularnewline
\hline
$\Omega_{ccc}^{++}\to\Omega_{cc}^+e^+\nu_e$&$4.95\pm0.92\pm2.12\pm1.16\pm0.10$&0.90&$1.23\pm0.23\pm0.53\pm0.29\pm0.02$\tabularnewline
\hline
$\Omega_{ccc}^{++}\to\Omega_{cc}^+\mu^+\nu_\mu$&$4.65\pm0.87\pm2.56\pm1.09\pm0.10$&0.82&$1.15\pm0.21\pm0.63\pm0.27\pm0.02$\tabularnewline
\hline
\end{tabular}
\end{table}
 
\begin{figure}[htbp!]
  	\begin{minipage}[t]{0.4\linewidth}
 	 \centering
  		\includegraphics[width=1.0\columnwidth]{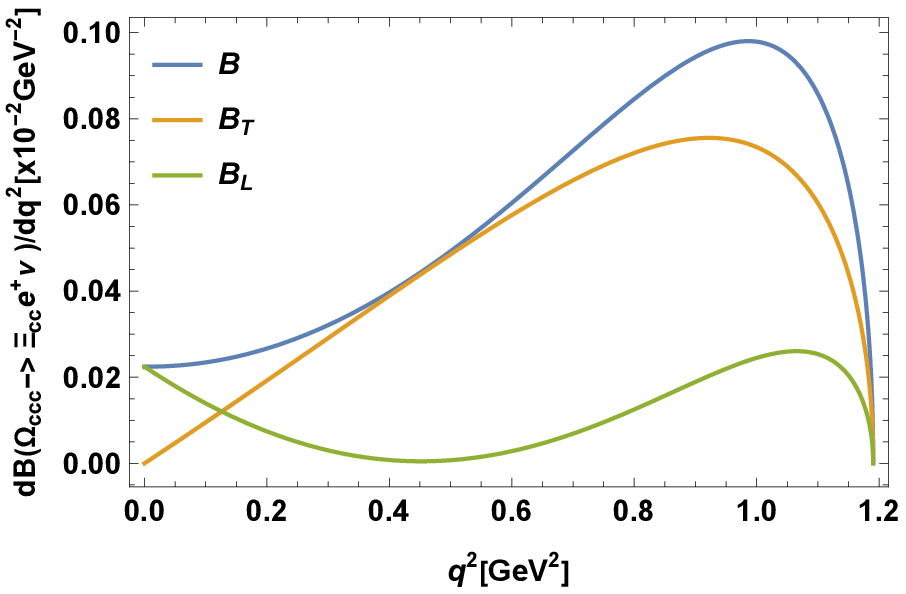}
   	 \end{minipage}
  	  \begin{minipage}[t]{0.4\linewidth}
  	\centering
  		\includegraphics[width=1.0\columnwidth]{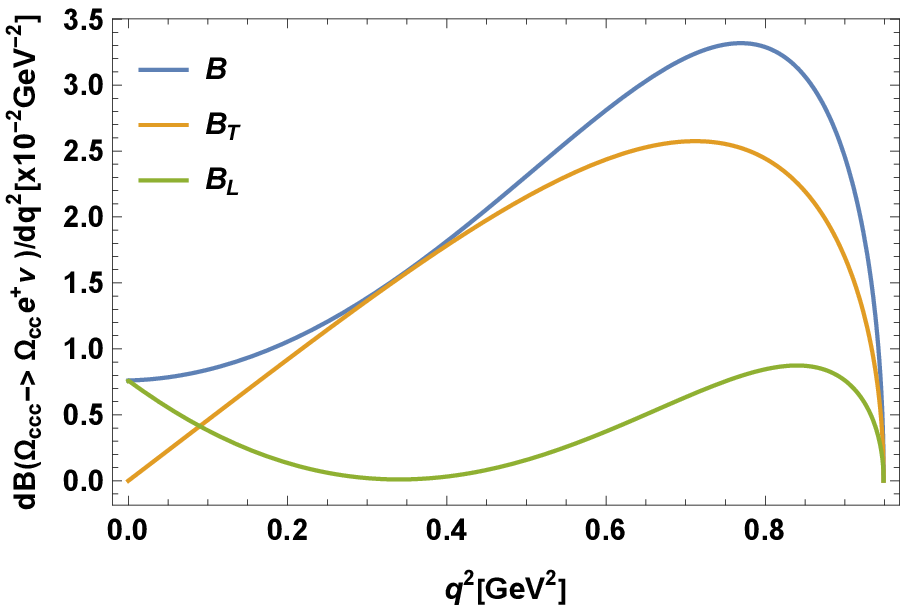}
    	\end{minipage}
 	\caption{ The differential branching fraction $dB/dq^2$ of spin-$3/2$ to spin-$1/2$ triply heavy baryon decay processes.  The $B_L$ and $B_T$ represent the contribution of longitudinal polarisation and transverse polarisation in branching fractions in Eq.~\eqref{gamma}.  } \label{dw}
\end{figure}

For the semi-leptonic decays, we can also study the $q^2$ distribution of branching fractions $(B,B_T,B_L)$ that correspond  to decay widths $(\Gamma,\Gamma_T,\Gamma_L)$, respectively.  
They are shown in Fig.~\ref{dw}. One can see that the longitudinal polarization contribution $B_L$ is smaller than transverse polarization contribution $B_T$ except in the region with $q^2< 0.3 {\rm GeV}^2$.  Since the the longitudinal polarization helicity amplitudes in Eq.~\eqref{ha} are proportional to the factor $1/\sqrt{q^2}$, the transverse polarization helicity amplitudes will be suppressed in small $q^2$ region comparing to longitudinal polarization helicity amplitudes.

\begin{figure}[htbp!]
	  \begin{minipage}[t]{0.4\linewidth}
	  \centering
		  \includegraphics[width=1.0\columnwidth]{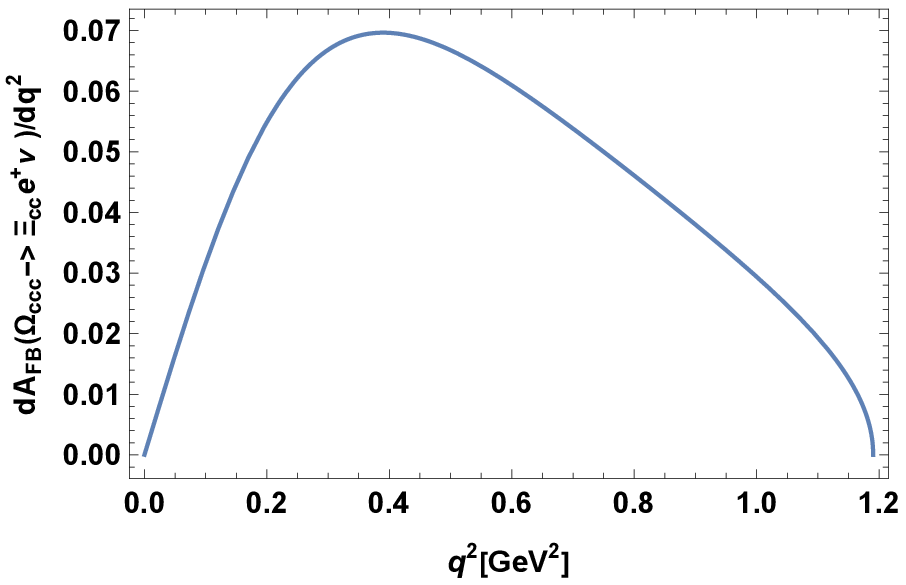}
   	 \end{minipage}
   	 \begin{minipage}[t]{0.4\linewidth}
	  \centering
		  \includegraphics[width=1.0\columnwidth]{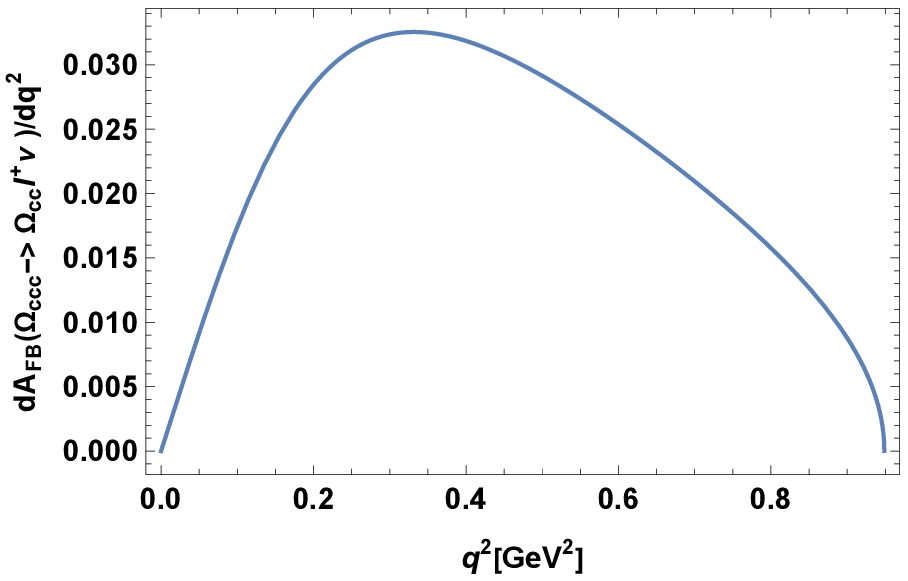}
	    \end{minipage}
	 \caption{ The $dA_{FB}/dq^2$ of spin-$3/2$ to spin-$1/2$ triply heavy baryon decay processes. } \label{AFB}
\end{figure}

To explore the $\theta$ distribution in Eq.~\eqref{angle}, one can define normalized the forward-backward asymmetry as
\begin{eqnarray}
	\frac{dA_{FB}}{dq^2}&=&\frac{\big[\int^1_0-\int^0_{-1}\big]d\cos\theta\frac{d^2\Gamma}{dq^2 d\cos\theta_\Lambda}}{d\Gamma/dq^2}=\frac{2L_2}{(4L_{1}-3L_{3})}.
\end{eqnarray}
Results for the normalized forward-backward asymmetry $A_{FB}$ are shown in Fig.\ref{AFB}. 

Recently, there is another analysis of the $\Omega_{ccc}^{++}$ decays in   light-front quark model~\cite{Zhao:2022vfr}. Their parametrization of form factors $f_1$, $f_4$, $g_1$ and $g_4$ is the same with Eq.~\eqref{eq:matrix_element_32VA} but we get the different value.  One can see that Ref.~\cite{Zhao:2022vfr} has adopted a different set of inputs for the shape parameters, and the differences in form factors caused  by the shape parameters can be treated as   systematic  uncertainties in this model. This problem can only be resolved  when the data is available in future.


\subsection{Nonleptonic decays}

For the color-allowed nonleptonic decay, the naive factorization  leads to: 
\begin{equation}
	i\mathcal{M}(\Omega_{ccc}^{++}\to\mathcal{B}_{cc}^{+}\;\mathcal{M})=\frac{G_{F}}{\sqrt{2}}V_{cq}^*V_{uq^\prime}a_1\langle \mathcal{B}_{cc}^{+}|\bar{q}\gamma_{\mu}(1-\gamma_5)c|\Omega_{ccc}^{++}\rangle \langle \mathcal{M}|\bar{q^\prime}\gamma^{\mu}(1-\gamma_5)u|0\rangle,\;a_1=C_1/N_c+C_2,\label{nonlep}
\end{equation}
where $N_c=3$ and $\mathcal{M}=\pi^+/K^+$ corresponding to the $q^\prime=\{d,s\}$ respectively.  The matrix element $\langle 0|\bar{q^\prime}\gamma_{\mu}(1-\gamma_5)u|\mathcal{M}\rangle$ can be expressed as
\begin{equation}
	\langle  \mathcal{M}|\bar{q^\prime}\gamma_{\mu}(1-\gamma_5)u| 0\rangle=if_M p_\mu,
\end{equation}
where $f_M$ is the decay constant of   $ {M}$. 
For the $\pi^+$ and $K^+$, we have $f_\pi=0.130\rm{GeV}$ and $f_K=0.155\rm{GeV}$\cite{Zyla:2020zbs}.

For the nonleptonic decays, one can also define the helicity amplitude as
 \begin{equation}
	HV_{\lambda^{\prime}}^\lambda \equiv\langle{\cal B}_{cc}^{+}(\lambda^{\prime})|\bar{q}\gamma_{\mu}Q_c|{\Omega}_{ccc}^{++}(\lambda)\rangle P_M^\mu\quad
{\rm and}\quad HA_{\lambda^{\prime}}^{\lambda}\equiv\langle{\cal B}_{cc}^{+}(\lambda^{\prime})|\bar{q}\gamma_{\mu}\gamma_{5}Q_c|{\Omega}_{ccc}^{++}(\lambda)\rangle P_M^\mu,
\end{equation}
where $P_M$ is the momentum of final meson state.
The expression of the helicity amplitudes can be given as
  \begin{eqnarray}
  &&HV_{\frac{1}{2}}^{\frac{1}{2}}=\frac{\sqrt{s_+}s_-m}{2\sqrt{6 q^2}M^2}\big[-2(M^\prime+M)\mathtt{f}_{1}+(M+M^{\prime})(\mathtt{f}_{2}+\mathtt{f}_{3})+\frac{m^2}{M-M^\prime}(\mathtt{f}_{2}-\mathtt{f}_{3})-2M\mathtt{f}_{4}\big],\notag\\
&&HA_{\frac{1}{2}}^{\frac{1}{2}}=-\frac{\sqrt{s_-}s_+m}{2\sqrt{6 q^2}M^2}\big[-2(M^\prime-M)\mathtt{g}_{1}+(M+M^{\prime})(\mathtt{g}_{2}+\mathtt{g}_{3})+\frac{m^2}{M-M^\prime}(\mathtt{g}_{2}-\mathtt{g}_{3})-2M\mathtt{g}_{4}\big],\notag\\
&&HV_{-\lambda^{\prime}}^{-\lambda}=-HV_{\lambda^{\prime}}^{\lambda},\quad HA_{-\lambda^{\prime},}^{-\lambda}=HA_{\lambda^{\prime}}^{\lambda}.
\end{eqnarray}
 Here $m$ is the mass of the emitted meson, and $q^2$ should be used as $m^2$ in the above equation. 
For the nonleptonic decays, the expression of  two body decay width is given as, 
\begin{eqnarray}
	&&\Gamma=\frac{G_F^2}{2}|V_{CKM}|^2a_1f_M^2\frac{\sqrt{s_+s_-}}{32\pi^2M^3}\bigg(|H_{-\frac{1}{2}}^{-\frac{1}{2}}|^2+|H_{\frac{1}{2}}^{\frac{1}{2}}|^2\bigg),
\end{eqnarray}
where $H_{\lambda^{\prime}}^{\lambda}=HV_{\lambda^{\prime}}^{\lambda}-HA_{\lambda^{\prime}}^{\lambda}$.
With the form factors one can obtain the results in Table~\ref{nonlepton}.

\begin{table}[!htb]
\caption{Numerical results of decay width and branching fraction in $\Omega_{ccc}^{++}$ nonleptonic decays. Uncertainties in branching fractions and decay widths are same with those in Table.~\ref{semilepton}.}
\label{nonlepton} %
\begin{tabular}{|c|c|c|c|c|c|c|c|c|}
\hline \hline
channel  & $\Gamma(\rm{GeV})$  & Branching fraction  \tabularnewline
\hline
$\Omega_{ccc}^{++}\to\Xi_{cc}^+K^+$&$(9.15\pm3.43\pm6.44\pm8.44\pm0.45)\times 10^{-17}$&$(2.27\pm0.85\pm1.60\pm2.10\pm0.11)\times 10^{-5}$\tabularnewline
\hline
$\Omega_{ccc}^{++}\to\Omega_{cc}^+K^+$&$(3.20\pm0.81\pm2.29\pm1.04\pm0.09)\times 10^{-15}$&$(7.95\pm2.00\pm5.69\pm2.58\pm0.21)\times 10^{-4}$\tabularnewline
\hline
$\Omega_{ccc}^{++}\to\Xi_{cc}^+\pi^+$&$(1.60\pm0.57\pm1.09\pm1.41\pm0.01)\times 10^{-15}$&$(3.97\pm1.41\pm2.71\pm3.50\pm0.01)\times 10^{-4}$\tabularnewline
\hline
$\Omega_{ccc}^{++}\to\Omega_{cc}^+\pi^+$&$(7.32\pm1.60\pm4.57\pm2.24\pm0.01)\times 10^{-14}$&$(1.82\pm0.40\pm1.13\pm0.56\pm0.00)\%$\tabularnewline
\hline
\multicolumn{3}{|c|}{Nonfactorizable contributions}\tabularnewline
\hline
channel  &  Branching fraction($N_c=2$) & Branching fraction($N_c=\infty$) \tabularnewline
\hline
$\Omega_{ccc}^{++}\to\Xi_{cc}^+K^+$&$1.94\times 10^{-5}$&$3.01\times 10^{-5}$\tabularnewline
\hline
$\Omega_{ccc}^{++}\to\Omega_{cc}^+K^+$&$6.79\times 10^{-4}$&$1.05\times 10^{-3}$\tabularnewline
\hline
$\Omega_{ccc}^{++}\to\Xi_{cc}^+\pi^+$&$3.39\times 10^{-4}$&$5.26\times 10^{-4}$\tabularnewline
\hline
$\Omega_{ccc}^{++}\to\Omega_{cc}^+\pi^+$&$1.55\%$&$2.41\%$\tabularnewline
\hline
\end{tabular}
\end{table}


\begin{figure}[htbp!] 
   \begin{minipage}[t]{0.4\linewidth}
  \centering
\includegraphics[width=1\columnwidth]{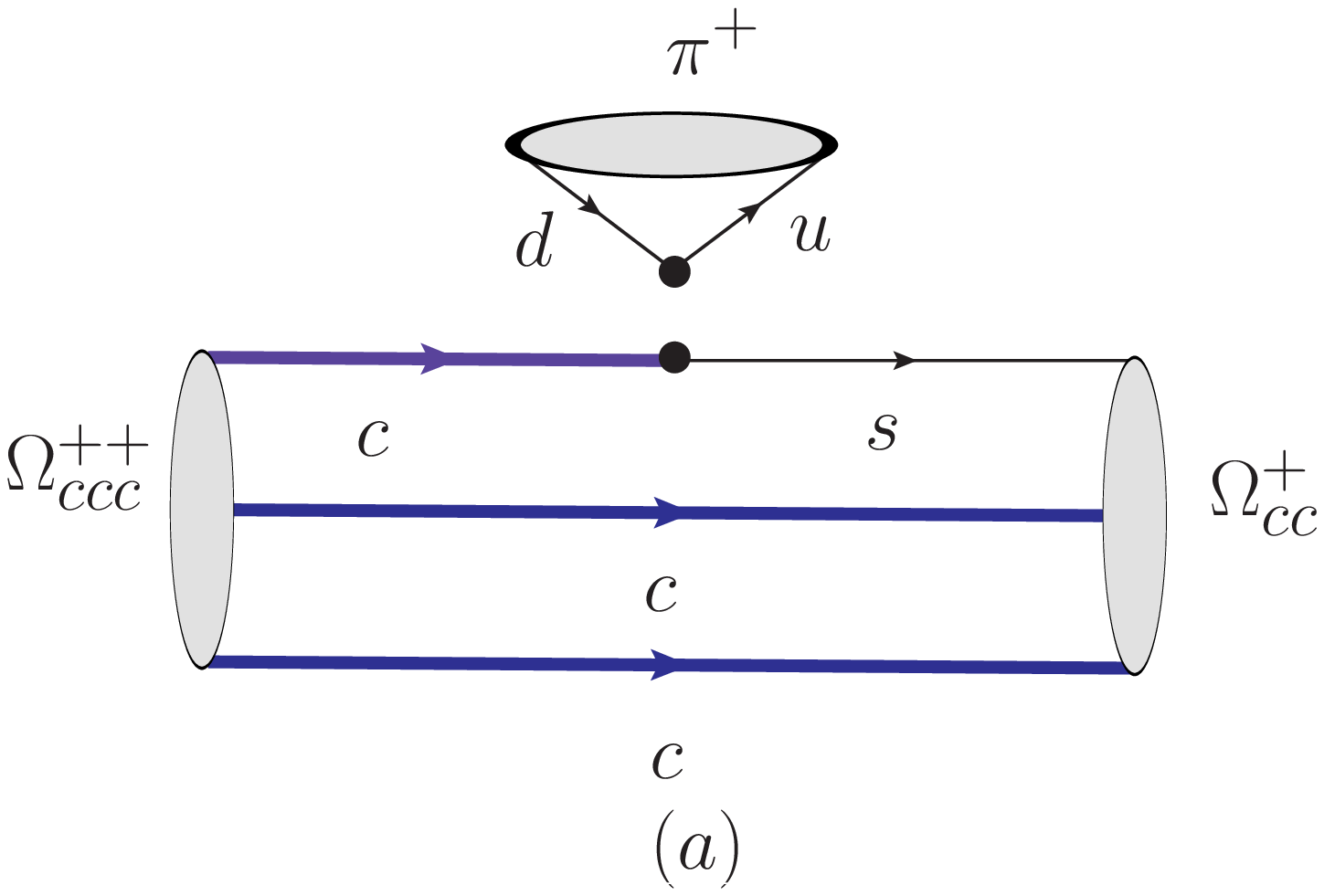}
 \end{minipage}
   \begin{minipage}[t]{0.4\linewidth}
  \centering
\includegraphics[width=1\columnwidth]{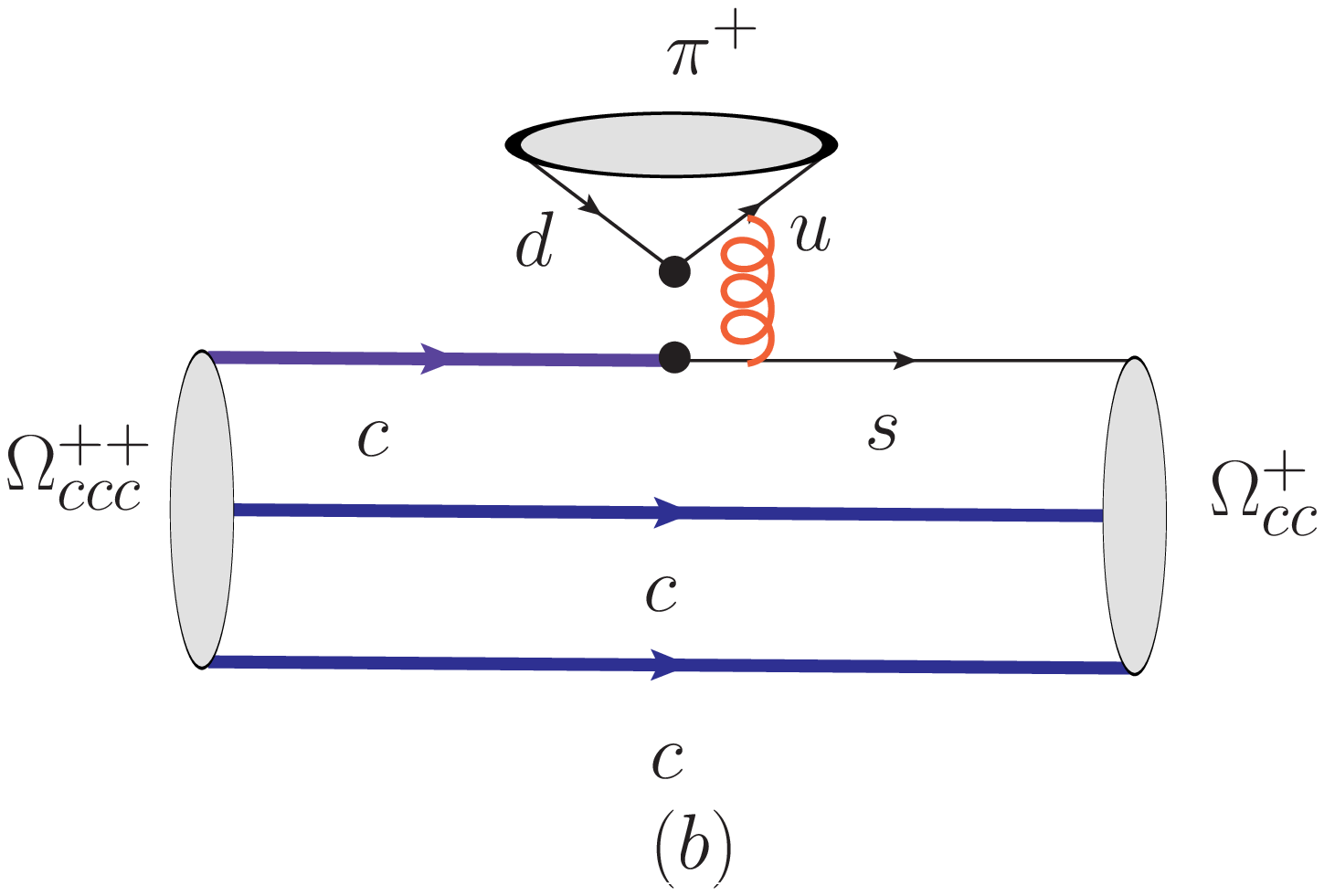}
 \end{minipage}
\caption{Factorizable (a) and non-factorizable (b) Feynman diagrams for   $\Omega_{ccc}^{++}\to\Omega_{cc}^+\pi^+$ decay. The black dots denote the effective operators. }
\label{fig:Feyn}
\end{figure} 

It is necessary to point out that the theoretical study of  nonleptonic decays is indeed  much more difficult than that of semileptonic decays, and one of the major obstacles in making reliable predictions is   non-factorizable contributions. In some nonlepotnic $B$ and $D$ decay modes such as color-suppressed channels, the  non-factorizable contributions are even predominant. However, for most color-allowed decays, the factorization approach can give a good estimate of  branching fractions, and thus in this work we have only focused on this type of decays. 

As an example,  we give a typical  Feynman diagram  for factorizable and nonfactorizable  contributions in $\Omega_{ccc}^{++}\to\Omega_{cc}^+\pi^+$ in Fig.~\ref{fig:Feyn}.  The factorizable contribution (in the left panel) is proportional to the Wilson coefficient $a_1= C_1/N_c+C_2$, while the nonfactorizable contribution  (in the right panel) is proportional to $C_1$ which is smaller than $a_1$.  In addition, there is also   suppression from the strong constant, though hard to estimate. Thus the nonfactorizable contributions are typically suppressed. 

In the literature, for example in Ref.~\cite{Ali:1998eb}, it has been suggested that the  nonfactorizable contributions in $B$ decays can be estimated through a variation of $N_c$ in Wilson coefficient $a_1= C_1/N_c+C_2$.  In this work, we also adopt this strategy  and give an estimate of the non-factorizable contributions by varying    $N_c$ to 2 and $\infty$. Results  are given in Tab.~\ref{nonlepton}, and ${\cal O}(30)\%$ corrections are found.

From the results in Tab.~\ref{nonlepton},, one can see that the $\Omega_{ccc}^{++}\to \Omega_{cc}^+\pi^+$ has a sizable branching fraction. Thus  it is presumable that an experimental search through this channel can lead to the discovery of  $\Omega_{ccc}^{++}$.

\subsection{SU(3) analysis}
Since the doubly charmed baryons and light mesons in final state compose the triplet and octet in SU(3) symmetry, the decay width of these four processes will satisfy the SU(3) relations:
\begin{eqnarray}
&&\Gamma(\Omega_{ccc}^{++}\to \Xi_{cc}^{+}\ell^+\nu_l)=\frac{|V_{cd}|^2}{|V_{cs}|^2}\Gamma(\Omega_{ccc}^{++}\to \Omega_{cc}^{+}\ell^+\nu),\notag\\
&&\Gamma(\Omega_{ccc}^{++}\to\Xi_{cc}^+K^+)=\frac{|V_{cd}V_{us}|^2}{|V_{cs}V_{ud}|^2}\Gamma(\Omega_{ccc}^{++}\to\Omega_{cc}^+\pi^+)=\frac{|V_{cd}|^2}{|V_{cs}|^2}\Gamma(\Omega_{ccc}^{++}\to\Omega_{cc}^+K^+)=\frac{|V_{us}|^2}{|V_{ud}|^2}\Gamma(\Omega_{ccc}^{++}\to\Xi_{cc}^+\pi^+),\label{su3}
\end{eqnarray}
where $V_{qq^\prime}$ are CKM matrix elements. 
Though the SU(3) analysis can not give the numerical prediction precisely,  the relations  in Eq.~\eqref{su3} can be tested with these numerical results. The ratio of decay width can be  estimated by LFQM and SU(3) symmetry simultaneously,
\begin{eqnarray}
	&&\bigg(\frac{\Gamma(\Omega_{ccc}^{++}\to\Xi_{cc}^+K^+)}{\Gamma(\Omega_{ccc}^{++}\to\Omega_{cc}^+\pi^+)}\bigg)_{SU(3)}=\frac{|V_{cd}V_{us}|^2}{|V_{cs}V_{ud}|^2}=0.0026,\notag\\
&&	\bigg(\frac{\Gamma(\Omega_{ccc}^{++}\to\Xi_{cc}^+K^+)}{\Gamma(\Omega_{ccc}^{++}\to\Omega_{cc}^+K^+)}\bigg)_{SU(3)}=
	\bigg(\frac{\Gamma(\Omega_{ccc}^{++}\to\Xi_{cc}^+\pi^+)}{\Gamma(\Omega_{ccc}^{++}\to\Omega_{cc}^+\pi^+)}\bigg)_{SU(3)}=\frac{|V_{cd}|^2}{|V_{cs}|^2}=0.050,\notag\\
	&&	\bigg(\frac{\Gamma(\Omega_{ccc}^{++}\to\Xi_{cc}^+K^+)}{\Gamma(\Omega_{ccc}^{++}\to\Xi_{cc}^+\pi^+)}\bigg)_{SU(3)}=
	\bigg(\frac{\Gamma(\Omega_{ccc}^{++}\to\Omega_{cc}^+K^+)}{\Gamma(\Omega_{ccc}^{++}\to\Omega_{cc}^+\pi^+)}\bigg)_{SU(3)}=\frac{|V_{us}|^2}{|V_{ud}|^2}=0.053,\notag\\
	&&\bigg(\frac{\Gamma(\Omega_{ccc}^{++}\to\Xi_{cc}^+\ell^+\nu_l)}{\Gamma(\Omega_{ccc}^{++}\to\Omega_{cc}^+\ell^+\nu_l)}\bigg)_{LFQM}=0.070,\quad	\bigg(\frac{\Gamma(\Omega_{ccc}^{++}\to\Xi_{cc}^+K^+)}{\Gamma(\Omega_{ccc}^{++}\to\Omega_{cc}^+\pi^+)}\bigg)_{LFQM}=0.0012,\notag\\
&&	\bigg(\frac{\Gamma(\Omega_{ccc}^{++}\to\Xi_{cc}^+K^+)}{\Gamma(\Omega_{ccc}^{++}\to\Omega_{cc}^+K^+)}\bigg)_{LFQM}=0.0285,\quad\bigg(\frac{\Gamma(\Omega_{ccc}^{++}\to\Xi_{cc}^+\pi^+)}{\Gamma(\Omega_{ccc}^{++}\to\Omega_{cc}^+\pi^+)}\bigg)_{LFQM}=0.0208,\notag\\
	&&	\bigg(\frac{\Gamma(\Omega_{ccc}^{++}\to\Xi_{cc}^+K^+)}{\Gamma(\Omega_{ccc}^{++}\to\Xi_{cc}^+\pi^+)}\bigg)_{LFQM}=0.060,\quad
	\bigg(\frac{\Gamma(\Omega_{ccc}^{++}\to\Omega_{cc}^+K^+)}{\Gamma(\Omega_{ccc}^{++}\to\Omega_{cc}^+\pi^+)}\bigg)_{LFQM}=0.044.
\end{eqnarray}

It can be seen that SU(3) flavor symmetry has a large breaking effect in nonleptonic processes.
The SU(3) symmetry breaking effects mostly arise from the discrepancy of form factors in Table.~\ref{ff}, since the form factors should the same in the SU(3) symmetry analysis.   Besides, the decay constants $f_K$ and $f_\pi$ can also contribute to the SU(3) symmetry breaking.
Compared to the branching ratios: $\Gamma(\Omega_{ccc}^{++}\to\Xi_{cc}^{+}\pi^+)/\Gamma(\Omega_{ccc}^{++}\to\Xi_{cc}^{+}K^+)$ and $\Gamma(\Omega_{ccc}^{++}\to\Omega_{cc}^{+}\pi^+)/\Gamma(\Omega_{ccc}^{++}\to\Omega_{cc}^{+}K^+)$ in  SU(3) and LFQM method, one can notice that since the momentum $q$ is on the mass shell of meson $\mathcal M$, the form factor and helicity amplitude will have very large SU(3) symmetry breaking in the two processes: $\Omega_{ccc}^{++}\to\Xi_{cc}^+K^+$ and $\Omega_{ccc}^{++}\to\Omega_{cc}^+\pi^+$. This analysis is consisted with our calculation in which the SU(3) symmetry breaking of these two processes is almost $50\%$.
Therefore, the mass difference of final mesons  and form factors will contribute to the SU(3) symmetry breaking effects together.

\section{Summary}

In summary, the spin-$3/2$ triply heavy baryon $\Omega_{ccc}^{++}$ and doubly heavy baryons $\Xi_{cc}^+$, $\Omega_{cc}^+$ state are calculated in the light-front quark model. 
For analyzing the baryons, we use the diquark picture in our calculation. In this picture, the diquark $[Q_cQ_c]$ can play a role like anti-quark so that the baryon formed by a quark and spectator diquark can be treated as meson.  	
Using the expression of hadron states in LFQM, we extract the form factor defined in hadron matrix element. 
For studying the $q^2$ dependence of the form factors, we have employed  the pole model. 

With the help of helicity amplitude, the differential and integrated decay width have been calculated.  
Using the lifetime of $\Omega_{ccc}^{++}$, we have predicted  branching fractions  of the semileptonic decays: $B(\Omega_{ccc}^{++}\to\Omega_{cc}^+e^+\nu_e)=1.23\%$ and $B(\Omega_{ccc}^{++}\to\Xi_{cc}^+e^+\nu_e)=0.08\%$.
The forward-backward asymmetry of angle $\theta$ is also studied. 
 Compared to with the SU(3) symmetry analysis, our results show a sizable  SU(3) symmetry break effects in nonleptonic processes and it mainly comes form the mass difference of final baryon and meson states. 
We have estimated   branching fractions for  a few  color-allowed processes and found that the $\Omega_{ccc}^{++}\to\Omega_{cc}^+\pi^+$ has a sizable branching fraction:  $\mathcal{B}(\Omega_{ccc}^{++}\to\Omega_{cc}^+\pi^+)=1.82\%$.   We point out that the $\Omega_{ccc}^{++}\to\Omega_{cc}^+\pi^+$ is a golden channel for the discovery of the $\Omega_{ccc}^{++}$.  This analysis can provide a useful  reference for future experimental and theoretical studies. 

\section*{Acknowledgements}

We thank Prof. Zhen-Xing Zhao for useful discussions. 
This work was supported in part by NSFC under Grant Nos.12147147, 11735010, 11905126, U2032102, 12061131006 and 12125503.

  \appendix
  \section{Helicity form factors}
  \label{hff}
  One can also parametrize the matrix element with the helicity form factors: 
  \begin{eqnarray}
&&\langle{\cal B}_{cc}^{+}(P^{\prime},S^{\prime}=\frac{1}{2},S_{z}^{\prime})|\bar{q}\gamma^{\mu}\gamma_{5}c|{\Omega}_{ccc}^{++}(P,S=\frac{3}{2},S_{z})\rangle\notag\\
&&=\bar u(p^\prime, s^\prime)\bigg(-g_0\frac{M}{s_+}\frac{(M-M^\prime)P^{\prime\lambda} q^\mu}{q^2}\notag   \\
&&\quad-g_+\frac{M}{s_-}\frac{(M+M^\prime)P^{\prime\lambda} (q^2(P^\mu+P^{\prime\mu})-q^\mu(M^2-M^{\prime2}))}{q^2 s_+}\notag\\
&&\quad+g_\perp\frac{M}{s_-}(P^{\prime\lambda}\gamma^\mu-\frac{2P^{\prime\lambda}(MP^{\prime\mu}+M^\prime P^\mu)}{s_+})\notag\\
&&\quad+g_{\perp\prime}\frac{M}{s_-}(P^{\prime\lambda}\gamma^\mu-\frac{2P^{\prime\lambda} P^{\mu}}{M}+\frac{2P^{\prime\lambda}(M P^{\prime\mu}+M^\prime P^\mu)}{s_+}+\frac{s_-g^{\lambda\mu}}{M})\bigg)u_\lambda(p, s),
\end{eqnarray}
\begin{eqnarray}
&&\langle{\cal B}_{cc}^{+}(P^{\prime},S^{\prime}=\frac{1}{2},S_{z}^{\prime})|\bar{q}\gamma^{\mu}c|{\Omega}_{ccc}^{++}(P,S=\frac{3}{2},S_{z})\rangle\notag\\
&&=\bar u(p^\prime, s^\prime)\bigg(f_0\frac{M}{s_-}\frac{(M+M^\prime)P^{\prime\lambda} q^\mu}{q^2}\notag\\
&&\quad+f_+\frac{M}{s_+}\frac{(M-M^\prime)P^{\prime\lambda} (q^2(P^\mu+P^{\prime\mu})-q^\mu(M^2-M^{\prime2}))}{q^2 s_-}\notag\\
&&\quad+f_\perp\frac{M}{s_+}(P^{\prime\lambda}\gamma^\mu+\frac{2P^{\prime\lambda}(MP^{\prime\mu}-M^\prime P^\mu)}{s_-})\notag\\
&&\quad+f_{\perp\prime}\frac{M}{s_+}(P^{\prime\lambda}\gamma^\mu+\frac{2P^{\prime\lambda} P^{\mu}}{M}-\frac{2P^{\prime\lambda}(M P^{\prime\mu}-M^\prime P^\mu)}{s_-}-\frac{s_+g^{\lambda\mu}}{M})\bigg)\gamma_5u_\lambda(p, s).
\end{eqnarray}
Then one can derive these form factors by applying the transformation to the form factors in Eq.~\eqref{eq:matrix_element_32VA} as
\begin{eqnarray}
\begin{pmatrix}
f_0\\
f_+\\
f_\perp\\
f^\prime_{\perp}
\end{pmatrix}=
\begin{pmatrix}
\frac{-s_-}{M^2}&\frac{s_-(M^2-M^{\prime2}+q^2)}{2M^2(M-M^\prime)(M+M^\prime)}
&\frac{s_-(M^2-M^{\prime2}-q^2)}{2M^2(M-M^\prime)(M+M^\prime)}&\frac{-s_-}{M(M+M^\prime)}\\
\frac{-s_+}{M^2}&\frac{s_+s_-}{2M^2(M-M^\prime)^2}&\frac{s_+s_-}{2M^2(M-M^\prime)^2}&-\frac{M^2-M^{\prime2}+q^2}{M(M-M^\prime)}\\
\frac{s_+}{M^2}&0&0&1\\
0&0&0&-1
\end{pmatrix}
\begin{pmatrix}
f_1\\
f_2\\
f_3\\
f_4
\end{pmatrix},\notag\\
\begin{pmatrix}
g_0\\
g_+\\
g_\perp\\
g^\prime_{\perp}
\end{pmatrix}=
\begin{pmatrix}
\frac{-s_+}{M^2}&\frac{-s_+(M^2-M^{\prime2}+q^2)}{2M^2(M-M^\prime)^2}
&\frac{-s_+(M^2-M^{\prime2}-q^2)}{2M^2(M-M^\prime)^2}&\frac{s_+}{M(M-M^\prime)}\\
\frac{-s_-}{M^2}&\frac{-s_+s_-}{2M^2(M-M^\prime)(M+M^\prime)}&\frac{-s_+s_-}{2M^2(M-M^\prime)(M+M^\prime)}&-\frac{M^2-M^{\prime2}+q^2}{M(M+M^\prime)}\\
\frac{s_-}{M^2}&0&0&-1\\
0&0&0&1
\end{pmatrix}
\begin{pmatrix}
g_1\\
g_2\\
g_3\\
g_4
\end{pmatrix}.
\end{eqnarray}
With the help of the definition of helicity form factors, the helicity amplitude can be given in a very simple form,
 \begin{eqnarray}
	&&HV_{\frac{1}{2},0}^{\frac{1}{2}}=HV_{-\frac{1}{2},0}^{-\frac{1}{2}}=f_+\frac{(M-M^\prime)\sqrt{s_-}}{\sqrt{6}\sqrt{q^2}},\quad HV_{\frac{1}{2},t}^{\frac{1}{2}}=HV_{-\frac{1}{2},t}^{-\frac{1}{2}}=f_0\frac{(M+M^\prime)\sqrt{s_+}}{\sqrt{6}\sqrt{q^2}},\notag\\
	&&HV_{-\frac{1}{2},-1}^{\frac{1}{2}}=-HV_{\frac{1}{2},1}^{-\frac{1}{2}}=f_\perp\frac{\sqrt{s_-}}{\sqrt{3}},\quad HV_{\frac{1}{2},-1}^{\frac{3}{2}}=-HV_{-\frac{1}{2},1}^{-\frac{3}{2}}=f^\prime_\perp\sqrt{s_-},
\end{eqnarray}
 and 
 \begin{eqnarray}
	&&HA_{\frac{1}{2},0}^{\frac{1}{2}}=HA_{-\frac{1}{2},0}^{-\frac{1}{2}}=g_+\frac{(M+M^\prime)\sqrt{s_+}}{\sqrt{6}\sqrt{q^2}},\quad HA_{\frac{1}{2},t}^{\frac{1}{2}}=HA_{-\frac{1}{2},t}^{-\frac{1}{2}}=g_0\frac{(M-M^\prime)\sqrt{s_-}}{\sqrt{6}\sqrt{q^2}},\notag\\
	&&HA_{-\frac{1}{2},-1}^{\frac{1}{2}}=HA_{\frac{1}{2},1}^{-\frac{1}{2}}=-g_\perp\frac{\sqrt{s_+}}{\sqrt{3}},\quad HA_{\frac{1}{2},-1}^{\frac{3}{2}}=HA_{-\frac{1}{2},1}^{-\frac{3}{2}}=g^\prime_\perp\sqrt{s_+}.
\end{eqnarray}
Using these results, one can find that each  helicity amplitude only has one form factor contribution. This form  can greatly simplify our analysis and  will be helpful for   further studies in  future.


\begin{thebibliography}{}
\bibitem{CDF:2003cab}
D.~Acosta \textit{et al.} [CDF],
Phys. Rev. Lett. \textbf{93}, 072001 (2004)
doi:10.1103/PhysRevLett.93.072001
[arXiv:hep-ex/0312021 [hep-ex]].

\bibitem{LHCb:2014zfx}
R.~Aaij \textit{et al.} [LHCb],
Phys. Rev. Lett. \textbf{112}, no.22, 222002 (2014)
doi:10.1103/PhysRevLett.112.222002
[arXiv:1404.1903 [hep-ex]].

\bibitem{LHCb:2019kea}
R.~Aaij \textit{et al.} [LHCb],
Phys. Rev. Lett. \textbf{122}, no.22, 222001 (2019)
doi:10.1103/PhysRevLett.122.222001
[arXiv:1904.03947 [hep-ex]].

\bibitem{LHCb:2017iph}
R.~Aaij \textit{et al.} [LHCb],
Phys. Rev. Lett. \textbf{119}, no.11, 112001 (2017)
doi:10.1103/PhysRevLett.119.112001
[arXiv:1707.01621 [hep-ex]].

\bibitem{Belle:2017ext}
J.~Yelton \textit{et al.} [Belle],
Phys. Rev. D \textbf{97}, no.5, 051102 (2018)
doi:10.1103/PhysRevD.97.051102
[arXiv:1711.07927 [hep-ex]].

\bibitem{LHCb:2018vuc}
R.~Aaij \textit{et al.} [LHCb],
Phys. Rev. Lett. \textbf{121}, no.7, 072002 (2018)
doi:10.1103/PhysRevLett.121.072002
[arXiv:1805.09418 [hep-ex]].

\bibitem{LHCb:2019bem}
R.~Aaij \textit{et al.} [LHCb],
Phys. Rev. Lett. \textbf{122}, no.23, 232001 (2019)
doi:10.1103/PhysRevLett.122.232001
[arXiv:1904.00081 [hep-ex]].

\bibitem{Hasenfratz:1980ka}
P.~Hasenfratz, R.~R.~Horgan, J.~Kuti and J.~M.~Richard,
Phys. Lett. B \textbf{94}, 401-404 (1980)
doi:10.1016/0370-2693(80)90906-5

\bibitem{Bjorken:1985ei}
J.~D.~Bjorken,
AIP Conf. Proc. \textbf{132}, 390-403 (1985)
doi:10.1063/1.35379

\bibitem{Silvestre-Brac:1996myf}
B.~Silvestre-Brac,
Few Body Syst. \textbf{20}, 1-25 (1996)
doi:10.1007/s006010050028

\bibitem{Vijande:2004at}
J.~Vijande, H.~Garcilazo, A.~Valcarce and F.~Fernandez,
Phys. Rev. D \textbf{70}, 054022 (2004)
doi:10.1103/PhysRevD.70.054022
[arXiv:hep-ph/0408274 [hep-ph]].

\bibitem{Roberts:2007ni}
W.~Roberts and M.~Pervin,
Int. J. Mod. Phys. A \textbf{23}, 2817-2860 (2008)
doi:10.1142/S0217751X08041219
[arXiv:0711.2492 [nucl-th]].

\bibitem{Martynenko:2007je}
A.~P.~Martynenko,
Phys. Lett. B \textbf{663}, 317-321 (2008)
doi:10.1016/j.physletb.2008.04.030
[arXiv:0708.2033 [hep-ph]].

\bibitem{Guo:2008he}
X.~H.~Guo, K.~W.~Wei and X.~H.~Wu,
Phys. Rev. D \textbf{78}, 056005 (2008)
doi:10.1103/PhysRevD.78.056005
[arXiv:0809.1702 [hep-ph]].

\bibitem{Yu:2017zst}
F.~S.~Yu, H.~Y.~Jiang, R.~H.~Li, C.~D.~L\"u, W.~Wang and Z.~X.~Zhao,
Chin. Phys. C \textbf{42}, no.5, 051001 (2018)
doi:10.1088/1674-1137/42/5/051001
[arXiv:1703.09086 [hep-ph]].

\bibitem{Qin:2021wyh}
Q.~Qin, Y.~J.~Shi, W.~Wang, G.~H.~Yang, F.~S.~Yu and R.~Zhu,
Phys. Rev. D \textbf{105}, no.3, L031902 (2022)
doi:10.1103/PhysRevD.105.L031902
[arXiv:2108.06716 [hep-ph]].

\bibitem{Hu:2019bqj}
X.~H.~Hu and Y.~J.~Shi,
Eur. Phys. J. C \textbf{80}, no.1, 56 (2020)
doi:10.1140/epjc/s10052-020-7635-1
[arXiv:1910.07909 [hep-ph]].

\bibitem{Han:2021gkl}
J.~J.~Han, R.~X.~Zhang, H.~Y.~Jiang, Z.~J.~Xiao and F.~S.~Yu,
Eur. Phys. J. C \textbf{81}, no.6, 539 (2021)
doi:10.1140/epjc/s10052-021-09239-w
[arXiv:2102.00961 [hep-ph]].

\bibitem{Yu:2019lxw}
F.~S.~Yu,
Sci. China Phys. Mech. Astron. \textbf{63}, no.2, 221065 (2020)
doi:10.1007/s11433-019-1483-0
[arXiv:1912.10253 [hep-ex]].

\bibitem{LHCb:2018pcs}
R.~Aaij \textit{et al.} [LHCb],
Phys. Rev. Lett. \textbf{121}, no.16, 162002 (2018)
doi:10.1103/PhysRevLett.121.162002
[arXiv:1807.01919 [hep-ex]].

\bibitem{LHCb:2018zpl}
R.~Aaij \textit{et al.} [LHCb],
Phys. Rev. Lett. \textbf{121}, no.5, 052002 (2018)
doi:10.1103/PhysRevLett.121.052002
[arXiv:1806.02744 [hep-ex]].

\bibitem{Cheng:2021qpd}
H.~Y.~Cheng,
[arXiv:2109.01216 [hep-ph]].

\bibitem{LHCb:2022rpd}
R.~Aaij \textit{et al.} [LHCb],
JHEP \textbf{05}, 038 (2022)
doi:10.1007/JHEP05(2022)038
[arXiv:2202.05648 [hep-ex]].

\bibitem{Brambilla:2005yk}
N.~Brambilla, A.~Vairo and T.~Rosch,
Phys. Rev. D \textbf{72}, 034021 (2005)
doi:10.1103/PhysRevD.72.034021
[arXiv:hep-ph/0506065 [hep-ph]].

\bibitem{Jia:2006gw}
Y.~Jia,
JHEP \textbf{10}, 073 (2006)
doi:10.1088/1126-6708/2006/10/073
[arXiv:hep-ph/0607290 [hep-ph]].

\bibitem{Zhang:2009re}
J.~R.~Zhang and M.~Q.~Huang,
Phys. Lett. B \textbf{674}, 28-35 (2009)
doi:10.1016/j.physletb.2009.02.056
[arXiv:0902.3297 [hep-ph]].

\bibitem{Brambilla:2009cd}
N.~Brambilla, J.~Ghiglieri and A.~Vairo,
Phys. Rev. D \textbf{81}, 054031 (2010)
doi:10.1103/PhysRevD.81.054031
[arXiv:0911.3541 [hep-ph]].

\bibitem{Meinel:2010pw}
S.~Meinel,
Phys. Rev. D \textbf{82}, 114514 (2010)
doi:10.1103/PhysRevD.82.114514
[arXiv:1008.3154 [hep-lat]].

\bibitem{Chen:2011mb}
Y.~Q.~Chen and S.~Z.~Wu,
JHEP \textbf{08}, 144 (2011)
[erratum: JHEP \textbf{09}, 089 (2011)]
doi:10.1007/JHEP08(2011)144
[arXiv:1106.0193 [hep-ph]].

\bibitem{Wang:2011ae}
Z.~G.~Wang,
Commun. Theor. Phys. \textbf{58}, 723-731 (2012)
doi:10.1088/0253-6102/58/5/17
[arXiv:1112.2274 [hep-ph]].

\bibitem{Flynn:2011gf}
J.~M.~Flynn, E.~Hernandez and J.~Nieves,
Phys. Rev. D \textbf{85}, 014012 (2012)
doi:10.1103/PhysRevD.85.014012
[arXiv:1110.2962 [hep-ph]].

\bibitem{Llanes-Estrada:2011gwu}
F.~J.~Llanes-Estrada, O.~I.~Pavlova and R.~Williams,
Eur. Phys. J. C \textbf{72}, 2019 (2012)
doi:10.1140/epjc/s10052-012-2019-9
[arXiv:1111.7087 [hep-ph]].

\bibitem{Aliev:2012tt}
T.~M.~Aliev, K.~Azizi and M.~Savci,
JHEP \textbf{04}, 042 (2013)
doi:10.1007/JHEP04(2013)042
[arXiv:1212.6065 [hep-ph]].

\bibitem{Wei:2015gsa}
K.~W.~Wei, B.~Chen and X.~H.~Guo,
Phys. Rev. D \textbf{92}, no.7, 076008 (2015)
doi:10.1103/PhysRevD.92.076008
[arXiv:1503.05184 [hep-ph]].

\bibitem{Yang:2019lsg}
G.~Yang, J.~Ping, P.~G.~Ortega and J.~Segovia,
Chin. Phys. C \textbf{44}, no.2, 023102 (2020)
doi:10.1088/1674-1137/44/2/023102
[arXiv:1904.10166 [hep-ph]].

\bibitem{Wang:2018utj}
W.~Wang and J.~Xu,
Phys. Rev. D \textbf{97}, no.9, 093007 (2018)
doi:10.1103/PhysRevD.97.093007
[arXiv:1803.01476 [hep-ph]].

\bibitem{Huang:2021jxt}
F.~Huang, J.~Xu and X.~R.~Zhang,
Eur. Phys. J. C \textbf{81}, no.11, 976 (2021)
doi:10.1140/epjc/s10052-021-09729-x
[arXiv:2107.13958 [hep-ph]].

\bibitem{Jaus:1999zv}
W.~Jaus,
Phys. Rev. D \textbf{60}, 054026 (1999)
doi:10.1103/PhysRevD.60.054026

\bibitem{Jaus:1989au}
W.~Jaus,
Phys. Rev. D \textbf{41}, 3394 (1990)
doi:10.1103/PhysRevD.41.3394

\bibitem{Jaus:1991cy}
W.~Jaus,
Phys. Rev. D \textbf{44}, 2851-2859 (1991)
doi:10.1103/PhysRevD.44.2851

\bibitem{Cheng:1996if}
H.~Y.~Cheng, C.~Y.~Cheung and C.~W.~Hwang,
Phys. Rev. D \textbf{55}, 1559-1577 (1997)
doi:10.1103/PhysRevD.55.1559
[arXiv:hep-ph/9607332 [hep-ph]].

\bibitem{Cheng:2003sm}
H.~Y.~Cheng, C.~K.~Chua and C.~W.~Hwang,
Phys. Rev. D \textbf{69}, 074025 (2004)
doi:10.1103/PhysRevD.69.074025
[arXiv:hep-ph/0310359 [hep-ph]].

\bibitem{Cheng:2004yj}
H.~Y.~Cheng and C.~K.~Chua,
Phys. Rev. D \textbf{69}, 094007 (2004)
[erratum: Phys. Rev. D \textbf{81}, 059901 (2010)]
doi:10.1103/PhysRevD.69.094007
[arXiv:hep-ph/0401141 [hep-ph]].

\bibitem{Ke:2007tg}
H.~W.~Ke, X.~Q.~Li and Z.~T.~Wei,
Phys. Rev. D \textbf{77}, 014020 (2008)
doi:10.1103/PhysRevD.77.014020
[arXiv:0710.1927 [hep-ph]].

\bibitem{Wei:2009np}
Z.~T.~Wei, H.~W.~Ke and X.~Q.~Li,
Phys. Rev. D \textbf{80}, 094016 (2009)
doi:10.1103/PhysRevD.80.094016
[arXiv:0909.0100 [hep-ph]].

\bibitem{Ke:2012wa}
H.~W.~Ke, X.~H.~Yuan, X.~Q.~Li, Z.~T.~Wei and Y.~X.~Zhang,
Phys. Rev. D \textbf{86}, 114005 (2012)
doi:10.1103/PhysRevD.86.114005
[arXiv:1207.3477 [hep-ph]].

\bibitem{Ke:2017eqo}
H.~W.~Ke, N.~Hao and X.~Q.~Li,
J. Phys. G \textbf{46}, no.11, 115003 (2019)
doi:10.1088/1361-6471/ab29a7
[arXiv:1711.02518 [hep-ph]].

\bibitem{Zhu:2018jet}
J.~Zhu, Z.~T.~Wei and H.~W.~Ke,
Phys. Rev. D \textbf{99}, no.5, 054020 (2019)
doi:10.1103/PhysRevD.99.054020
[arXiv:1803.01297 [hep-ph]].

\bibitem{Hu:2020mxk}
X.~H.~Hu, R.~H.~Li and Z.~P.~Xing,
Eur. Phys. J. C \textbf{80}, no.4, 320 (2020)
doi:10.1140/epjc/s10052-020-7851-8
[arXiv:2001.06375 [hep-ph]].

\bibitem{Chang:2020wvs}
Q.~Chang, X.~L.~Wang and L.~T.~Wang,
Chin. Phys. C \textbf{44}, no.8, 083105 (2020)
doi:10.1088/1674-1137/44/8/083105
[arXiv:2003.10833 [hep-ph]].

\bibitem{Chen:2021ywv}
L.~Chen, Y.~W.~Ren, L.~T.~Wang and Q.~Chang,
Eur. Phys. J. C \textbf{82}, no.5, 451 (2022)
doi:10.1140/epjc/s10052-022-10391-0
[arXiv:2112.08016 [hep-ph]].

\bibitem{Ke:2019smy}
H.~W.~Ke, N.~Hao and X.~Q.~Li,
Eur. Phys. J. C \textbf{79}, no.6, 540 (2019)
doi:10.1140/epjc/s10052-019-7048-1
[arXiv:1904.05705 [hep-ph]].

\bibitem{Buchalla:1995vs}
G.~Buchalla, A.~J.~Buras and M.~E.~Lautenbacher,
Rev. Mod. Phys. \textbf{68}, 1125-1144 (1996)
doi:10.1103/RevModPhys.68.1125
[arXiv:hep-ph/9512380 [hep-ph]].

\bibitem{Zyla:2020zbs}
P.~A.~Zyla \textit{et al.} [Particle Data Group],
PTEP \textbf{2020}, no.8, 083C01 (2020)
doi:10.1093/ptep/ptaa104

\bibitem{Chua:2018lfa}
C.~K.~Chua,
Phys. Rev. D \textbf{99}, no.1, 014023 (2019)
doi:10.1103/PhysRevD.99.014023
[arXiv:1811.09265 [hep-ph]].

\bibitem{Shi:2019hbf}
Y.~J.~Shi, W.~Wang and Z.~X.~Zhao,
Eur. Phys. J. C \textbf{80}, no.6, 568 (2020)
doi:10.1140/epjc/s10052-020-8096-2
[arXiv:1902.01092 [hep-ph]].

\bibitem{Chua:2019yqh}
C.~K.~Chua,
Phys. Rev. D \textbf{100}, no.3, 034025 (2019)
doi:10.1103/PhysRevD.100.034025
[arXiv:1905.00153 [hep-ph]].

\bibitem{Li:2010bb}
G.~Li, F.~l.~Shao and W.~Wang,
Phys. Rev. D \textbf{82}, 094031 (2010)
doi:10.1103/PhysRevD.82.094031
[arXiv:1008.3696 [hep-ph]].

\bibitem{Verma:2011yw}
R.~C.~Verma,
J. Phys. G \textbf{39}, 025005 (2012)
doi:10.1088/0954-3899/39/2/025005
[arXiv:1103.2973 [hep-ph]].

\bibitem{Shi:2016gqt}
Y.~J.~Shi, W.~Wang and Z.~X.~Zhao,
Eur. Phys. J. C \textbf{76}, no.10, 555 (2016)
doi:10.1140/epjc/s10052-016-4405-1
[arXiv:1607.00622 [hep-ph]].

\bibitem{Faustov:2021qqf}
R.~N.~Faustov and V.~O.~Galkin,
Phys. Rev. D \textbf{105}, no.1, 014013 (2022)
doi:10.1103/PhysRevD.105.014013
[arXiv:2111.07702 [hep-ph]].

\bibitem{Zhao:2018mrg}
Z.~X.~Zhao,
Eur. Phys. J. C \textbf{78}, no.9, 756 (2018)
doi:10.1140/epjc/s10052-018-6213-2
[arXiv:1805.10878 [hep-ph]].

\bibitem{Zhao:2022vfr}
Z.~X.~Zhao,
[arXiv:2204.00759 [hep-ph]].

\bibitem{Ali:1998eb}
A.~Ali, G.~Kramer and C.~D.~Lu,
Phys. Rev. D \textbf{58}, 094009 (1998)
doi:10.1103/PhysRevD.58.094009
[arXiv:hep-ph/9804363 [hep-ph]].
  \end{thebibliography}
\end{document}